\documentclass[pra,floatfix,twocolumn,amsfonts,amsmath,amssymb,nofootinbib,reprint,superscriptaddress]{revtex4-2}
\pdfoutput=1
\usepackage{calligra}
 
\usepackage{graphicx}
\usepackage{dcolumn}
\usepackage{bm}
\usepackage[utf8]{inputenc}
\usepackage{booktabs}

\usepackage[colorlinks]{hyperref}
\usepackage[normalem]{ulem}
\usepackage{comment}
\usepackage{xcolor}
\usepackage{amsthm}
\usepackage{mathtools}
\usepackage{bbm}
\usepackage{bm}
\usepackage{graphicx}
\usepackage{enumerate}
\usepackage{tikz}
\usetikzlibrary{arrows}
\usepackage{soul}
\usepackage{csquotes}

\newtheorem{theorem}{Theorem}

\newtheorem{corollary}[]{Corollary}

\newtheorem{lemma}{Lemma}

\usepackage{microtype}

\newtheorem{definition}{Definition}

\def\bra#1{\mathinner{\langle{#1}|}}
\def\ket#1{\mathinner{|{#1}\rangle}}
\def\braket#1#2{\mathinner{\langle{#1|#2}\rangle}}
\def\ketbra#1#2{\mathinner{|{#1}\rangle\!\langle{#2}|}}

\def\a{\bm{a}}
\def\x{\bm{x}}
\def\w{\bm{w}}
\def\BraVert{\egroup\,\mid@vertical\,\bgroup}

\DeclareMathOperator{\Tr}{Tr}

\renewcommand\L{\mathcal{L}}

\newcommand{\HS}{\mathcal{H}}

\newcommand{\A}{\mathcal{A}}

\newcommand{\X}{\mathcal{X}}

\newcommand{\I}{\mathcal{I}}

\newcommand{\id}{\mathbbm{1}}

\renewcommand{\a}{\bm{a}}

\newcommand{\hippocom}[1]{\textit{\small\textcolor{cyan}{[H: #1]}}}

\newcommand{\andreas}[1]{{\color{green} #1}}
\newcommand{\andreascom}[1]{\textit{\small\textcolor{green}{[A: #1]}}}

\renewcommand{\S}{\mathcal{S}}

\begin{document}

\preprint{APS/123-QED}

\title{Paradox-free classical non-causality and unambiguous non-locality without entanglement are equivalent}

\author{Hippolyte Dourdent}
 \email{hippolyte.dourdent@uni-muenster.de}
\affiliation{ICFO-Institut de Ciencies Fotoniques, The Barcelona Institute of Science and Technology,\\ 08860 Castelldefels, Barcelona, Spain}\affiliation{Department for Quantum Technology, Universität Münster,
Heisenbergstraße 11, 48149 Münster, Germany}

\author{{Kyrylo Simonov}}
\affiliation{Fakult\"at f\"ur Mathematik, Universit\"at Wien, Oskar-Morgenstern-Platz 1, 1090 Vienna, Austria}

\author{{Andreas Leitherer}}
\affiliation{ICFO-Institut de Ciencies Fotoniques, The Barcelona Institute of Science and Technology,\\ 08860 Castelldefels, Barcelona, Spain}

\author{{Emanuel-Cristian Boghiu}}
\affiliation{Fakult\"at f\"ur Mathematik, Universit\"at Wien, Oskar-Morgenstern-Platz 1, 1090 Vienna, Austria}

\author{{Ravi Kunjwal}}
\affiliation{Universit\'e libre de Bruxelles, QuIC, Brussels, Belgium}
\affiliation{Aix-Marseille University, CNRS, LIS, Marseille, France}

\author{{Saronath Halder}}
\affiliation{Center for Theoretical Physics, Polish Academy of Sciences, Aleja Lotników 32/46, 02-668 Warsaw, Poland}
\affiliation{Department of Physics, School of Advanced Sciences, VIT-AP University, Amaravati 522241, Andhra Pradesh, India}

\author{Remigiusz Augusiak}
\affiliation{Center for Theoretical Physics, Polish Academy of Sciences, Aleja Lotników 32/46, 02-668 Warsaw, Poland}

\author{Antonio Acín}
\affiliation{ICFO-Institut de Ciencies Fotoniques, The Barcelona Institute of Science and Technology,\\ 08860 Castelldefels, Barcelona, Spain}
\affiliation{ICREA, Passeig Lluis Companys 23, 08010 Barcelona, Spain}

\date{\today}

\begin{abstract}
Closed timelike curves (CTCs) challenge our conception of causality by allowing information to loop back into its own past. Any consistent description of such scenarios must avoid time-travel paradoxes while respecting the no-new-physics principle, which requires that the set of operations available within any local spacetime region remain unchanged, irrespective of whether CTCs exist elsewhere. Within an information-theoretic framework, this leads to process functions: deterministic classical communication structures that remain logically consistent under arbitrary local operations, yet can exhibit correlations incompatible with any definite causal order\textemdash a phenomenon known as non-causality. In this work, we establish a correspondence between process functions and unambiguous complete product bases, i.e.~product bases in which every local state belongs to a unique local basis. This equivalence implies that non-causality of process functions is exactly mirrored by quantum nonlocality without entanglement (QNLWE)\textemdash the impossibility of perfectly distinguishing  separable states using local operations and causal classical communication\textemdash for such bases. Our results generalize previous special cases to arbitrary local dimensions and any number of parties, enable systematic constructions of non-causal process functions and unambiguous QNLWE bases, and implies, inter alia, that bipartite unambiguous QNLWE bases do not exist. Finally, this work reveals an unexpected connection between causal and non-signaling inequalities: every process function both maximally violates an associated causal inequality and yields a corresponding Bell inequality that admits no violation.

\end{abstract}

\maketitle

\medskip

\section{Introduction}
Although it remains unknown whether closed timelike curves (CTCs)\textemdash trajectories in spacetime that loop back into the past~\cite{lanczos1924stationare, godel49,luminet21}\textemdash exist in Nature, their logical possibility challenges our understanding of physical dynamics and  computer science~\cite{brun2003computers, bennett2009can, aaronson2009closed, vairogs2022quantum}.
CTCs are notorious for giving rise to time-travel paradoxes~\cite{baumeler21, smeenk_wuethrich, allen2014treating, baumeler2018computational, pienaar2013causality, aaronson201310} such as the grandfather antinomy, in which an effect prevents its own cause, and the information or bootstrap antinomy, where an effect becomes its own cause. At their core lies a tension between events that have already occurred and the freedom of agents to influence them, a tension that has fueled a long-standing debate across physics and philosophy. Stephen Hawking famously argued that time-travel should be impossible, proposing the so-called chronology protection conjecture to ``make the universe safe for historians''~\cite{hawking}. More nuanced accounts were offered by philosophers such as David Lewis, who, within his modal realist framework, argued that consistency can be restored through the Leibnizian notion of ``compossible facts'', i.e.~possible facts that can coexist with one another: what an agent can do relative to one set of facts may be impossible relative to another, more comprehensive, set that includes further facts~\cite{lewis}.  Igor Novikov formulated a self-consistency principle~\cite{novikov, novikov1989analysis}, according to which ``events on a closed timelike curve are already guaranteed to be self-consistent''. In this view, a time traveler retains the freedom to act, but her possible choices are constrained by a kind of ``fine-tuning'' or ``time-policing'' principle that prevents paradoxical outcomes.

Analyses such as the billiard-ball-through-a-wormhole scenario~\cite{echeverria1991billiard} demonstrated that apparently inconsistent setups can nonetheless admit self-consistent solutions. This led to the formulation of the ``no new physics'' principle~\cite{novikov}, which asserts that the physical laws (in particular, the set of possible physical operations) within any local spacetime region remain unchanged, regardless of whether CTCs exist elsewhere~\cite{baumeler}. Building on this idea, several consistent models have been proposed to investigate the interaction of quantum systems with operational, information-theoretic analogs of spacetime CTCs. Two prominent examples are the Deutsch-CTC (D-CTC) model~\cite{deutsch1991quantum}, motivated by the Everett interpretation, and the Post-selected-CTC (P-CTC) model~\cite{politzer, bennett, svetlichny}, inspired by quantum teleportation and based on post-selecting a Bell effect to implement a quantum channel to the past. While these frameworks successfully avoid causal paradoxes, this is achieved at the cost of abandoning either the ``no new physics'' principle or the assumption of freedom of choice, that is, the ability of agents to perform arbitrary operations in locally well-defined regions. Nevertheless, an alternative approach rooted in the study of quantum causality offers a framework free from these pathologies.\\

The notion of indefinite causal order \cite{costa26} originates from the  causaloid formalism~\cite{hardy05, hardy07}, which aims to bridge the dynamical causal structure of general relativity with the indeterminacy of quantum theory. This work questioned the fundamental nature of events taking place in a well-defined causal order and led to the development of the process matrix formalism~\cite{oreshkov1}. This framework demonstrates that indefinite causal orders  can emerge in scenarios involving spatially separated parties that  control laboratories in which quantum theory holds locally\textemdash they are allowed to perform any local quantum operations\textemdash while no assumptions on a fixed global spacetime structure are made. In this information-theoretic context, local events are defined by classical outcomes of quantum operations in the laboratories.  
Causal relations between two events are then defined as one-way non-signaling constraints that encapsulate the order of operations. (For analysis and discussions on the relation between information-theoretic and space-time causal structures, cf.~Ref~\cite{vilasini24}. For a brief overview of the debate surrounding the experimental realization of indefinite causal order, cf.~Refs~\cite{rozema24} and~\cite[Appendix G.1]{steffinlongo25}.). 

Process matrices preserve local physics while enabling globally acting processes that can violate causal inequalities~\cite{oreshkov1, branciard1}, a phenomenon termed \emph{non-causality}. These matrices can be seen as a linear subset of P-CTCs~\cite{araujo3} or pre-post-selected quantum states~\cite{silva17}. The deterministic classical counterpart of process matrices, namely \emph{process functions (PF)}~\cite{baumeler2, baumeler16, baumeler19}, was introduced to model classical communication scenarios beyond the standard causal ones. Various non-causal process functions for $n \geq 3$ parties have been identified~\cite{baumeler14, baumeler2, af, baumeler22, tobar, mejdoub25}. This is remarkable: despite lacking a well-defined causal order, these new classical communication structures preserve logical consistency, i.e.~avoid time travel paradoxes, while maintaining freedom of choice~\cite{baumeler, baumeler21}. 

Local Operations with Process Function (LOPF) can be used as resources, providing advantages over traditional Local Operations with Classical Communication (LOCC)~\cite{chitambar14}, while still preserving freedom of choice\textemdash unlike the broader class of LOCC without causal order (LOCC*)~\cite{akibue17}. In Ref.~\cite{kunjwal23a}, Kunjwal and Baumeler demonstrated that the SHIFT basis 
\begin{align}
\{\ket{000}\!, \ket{{+}01}\!, \ket{01{+}}\!, \ket{01{-}}\!,  \ket{1{+}0}\!, \ket{{-}01}\!, \ket{1{-}0}\!, \ket{111}\},
\label{eq:shift}
\end{align}
which cannot be measured locally under LOCC\textemdash a phenomenon known as \emph{``quantum non-locality without entanglement" (QNLWE)}~\cite{bennett_99, groisman01, niset06, Halder19, zhen_22}\textemdash is \textit{equivalent} to a tripartite non-causal process function, the \textit{Lugano (AF/BW) process}~\cite{af, baumeler2}. In this process, each party's input depends non-trivially on the outputs prepared by the other two parties, where all parties are both in each other's past and future, with no restrictions on the local operations they perform:
\begin{align}
    x_1 := a_3(a_2 \oplus 1), \quad x_2 := a_1(a_3 \oplus 1), \quad x_3 := a_2(a_1 \oplus 1),
    \label{eq:lugano}
\end{align}
where $x_1, x_2, x_3$ and $a_1, a_2, a_3$ label the input and output bits for the first, second, and third party, respectively. 

By equivalence we refer to the fact that local operations communicating via the Lugano process can implement the SHIFT measurement (the projective measurement in the basis shown in Eq.~\eqref{eq:shift}) and, conversely, the SHIFT measurement enables the realization of the classical channel associated to the Lugano process in a quantum circuit. Furthermore, all non-causal Boolean (i.e.~multi-bit) process functions without a global past can systematically transform into QNLWE multi-qubit bases, effectively trading definite causal order (causality of correlations) for local measurability~\cite{kunjwal23a}. \\

Several open questions arise from these results. Can every QNLWE multi-qubit basis be mapped onto a Boolean non-causal process function? Does this correspondence persist for higher-dimensional systems?  Another QNLWE canonical example\textemdash the two-qutrit domino basis~\cite{bennett_99}\textemdash appears to suggest a negative answer, as it would correspond to a bipartite process while logically consistent classical non-causality arises only for three parties or more, i.e.~all bipartite process function are causal~\cite{baumeler19}. In fact, the domino basis was shown to be perfectly measurable by LOCC*~\cite{akibue17}. QNLWE also emerges from unextendible product bases (UPBs), such as the SHIFT UPB $\{\ket{{+}01}\!, \ket{01{+}}\!, \ket{1{+}0}\!, \ket{{-}{-}{-}}\!\}$~\cite{bennett99,divincenzo03}, which can be perfectly discriminated by the global causal loop $(x_1:=a_3, x_2:=a_1,x_3:=a_2)$. However, this structure does not correspond to a valid process function, as it can trivially lead to paradoxes under certain local operations (cf.~Section \ref{sec:characterisation_pfs} for more details). These examples suggest that non-causal process functions may only be equivalent to a sub-class of QNLWE bases.

\section{Results} 

In this work, we demonstrate that logical consistency for deterministic classical communication resources under free local operations\textemdash which fully characterizes process functions\textemdash is equivalent to the assumption we term \textit{unambiguity} for \textit{complete} product bases and their associated separable projective measurements (Theorems \ref{thm:unamb_to_pf} and  \ref{thm:pftob}).
Using a unique fixed point characterisation (Section \ref{sec:characterisation_pfs}), as well as a  characterisation of (non-)causal process functions (Section~\ref{sec:causalnon-causal}), we demonstrate that $(i)$ every unambiguous product basis can be used to construct a process function (Section~\ref{sec:upbpf}), $(ii)$ every process function can be encoded in an unambiguous product basis (Section~\ref{sec:pfupb}), and $(iii)$ non-causality and QNLWE are equivalent in this context (Section~\ref{sec:ncqnlwe}). We illustrate the utility of this equivalence by constructing new non-causal process functions and QNLWE bases, thereby revealing its potential to uncover further results. In particular, it implies the non-existence of bipartite unambiguous QNLWE bases in any dimension (Section \ref{sec:noqnlwe}) as a direct consequence of the non-existence of bipartite non-causal process functions. We also remark that $(i)$ is closely related to a construction of Bell inequalities without violations~\cite{augusiak11,augusiak12}, uncovering an unexpected link between some non-signaling and causal inequalities (see Section~\ref{sec:upb}).

\section{Process functions}\label{sec:pf}
 Before presenting our main results, we introduce the framework of process functions, present their characterization in terms of fixed point uniqueness (Section \ref{sec:characterisation_pfs}),  and introduce a characterization for (non-)causal process functions (Section \ref{sec:causalnon-causal}).
 
 \medskip
 
\begin{figure}[ht!]
	\begin{center}
\includegraphics[width=0.9\columnwidth]{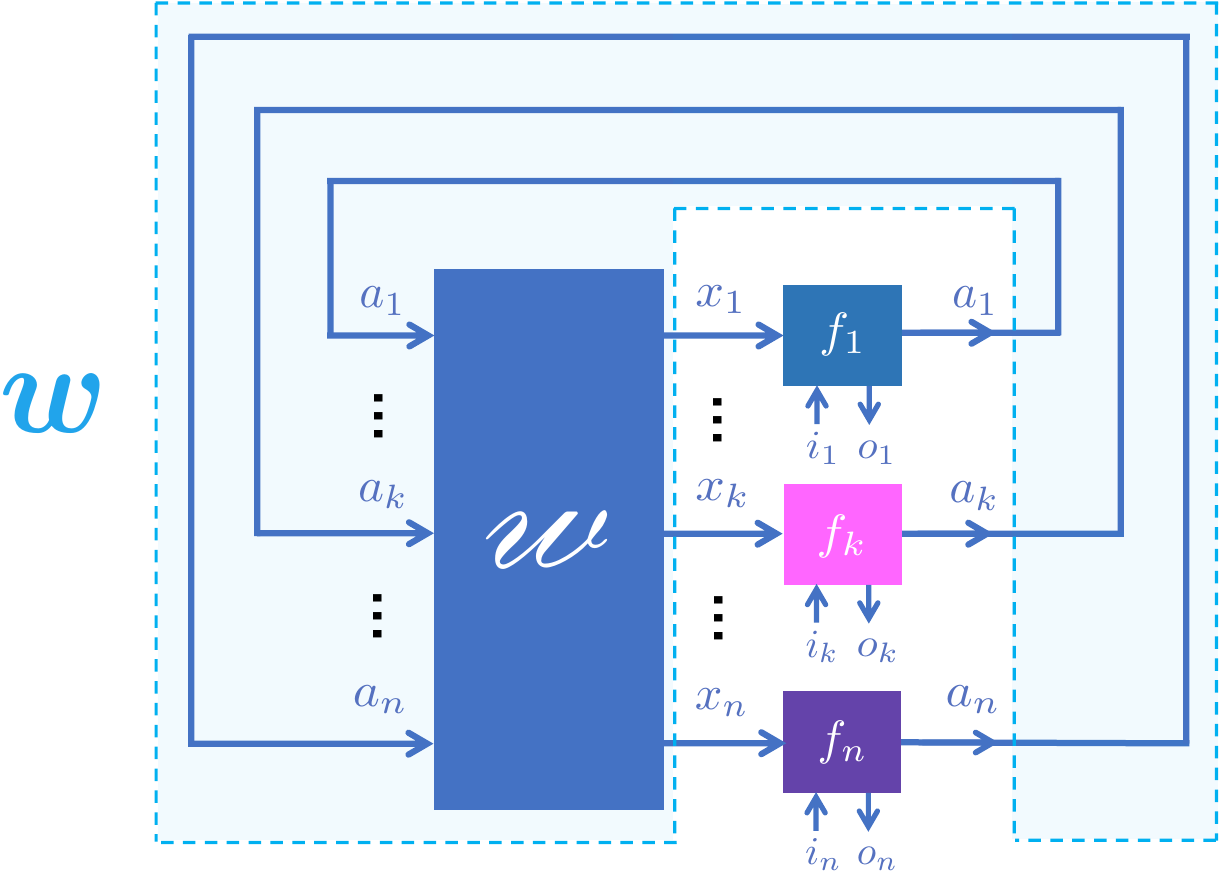}
	\end{center}
	\caption{In the process function scenario, 
    $n$ isolated parties each can perform free local interventions $f_k$ while conducting single-round communication that is modeled by the process function $\w$, which combines the classical channel $\text{\Large{\calligra w }}$  and retro-causal identity channels. Each party $k$  receives an external input $i_k$ and an incoming variable $x_k$ from $\w$. For a given $f_k$, each  party then produces the external output $o_k$ as well as an output $a_k$ which is  fed back into $\w$.} 
	\label{fig_pf}
\end{figure}

 Consider an $n-$partite scenario (cf.~Fig.~\ref{fig_pf}), where each party, labeled with a distinct $k\in\{1,...,n\}$, receives  classical variables $x_k\in\mathcal{X}_k$ and $i_k\in\mathcal{I}_k$ as inputs, and produces outputs $a_k\in\mathcal{A}_k$ and $o_k\in\mathcal{O}_k$. Here,\footnote{Our notation is distinct from that adopted in Refs.~\cite{baumeler2,baumeler19,tobar,kunjwal23a,kunjwal23}.} $\mathcal{X}_k$ and $\mathcal{A}_k$ represent the information exchanged with a shared communication resource and play a role analogous to hidden variables in Bell non-locality, mediating the correlations among the parties. In contrast, $\mathcal{I}_k$ and $\mathcal{O}_k$ correspond to the locally accessible input and output registers, analogous to the measurement settings and outcomes that determine the observed correlations in a Bell experiment. For compactness, we will use the notations $\mathcal{X}:=\bigtimes_{k=1}^n \mathcal{X}_k$, $\mathcal{A}:=\bigtimes_{k=1}^n \mathcal{A}_k$,   $\bm{x}=(x_1,...,x_n)$, 
$\bm{x}_{\backslash i}:=(x_1,...,x_{i-1},x_{i+1},..,x_n)$, $\bm{a}=(a_1,...,a_n)$, etc. The parties remain operationally isolated: they do not communicate directly but only interact through the shared communication resource. Their inputs $\mathcal{I}_k$ can be freely chosen, that is, they are not correlated with the shared resource. Each party executes a chosen classical 
operation modeled by a local stochastic channel $P(a_k,o_k|x_k,i_k)$ 
once and only once, i.e.~we consider single-round communication scenarios throughout this article.  

The observed correlations are
\begin{equation}
    P(\bm{o}|\bm{i}) = \sum_{\bm{a},\bm{x}} \Bigl(\prod_{k=1}^n P(a_k,o_k| x_k,i_k)\Bigr) P(\bm{x}| \bm{a}), \label{eq:correlation2}
\end{equation}
where $P(\bm{x}| \bm{a})$ denotes the (possibly stochastic) \emph{classical process} implemented by the shared communication resource. A classical process is \emph{logically consistent} if Eq.~\eqref{eq:correlation2} yields a valid conditional distribution $P(\bm{o}|\bm{i})$ for \emph{every} choice of local operations, i.e.~$P(\bm{o}|\bm{i}) \geq 0$ and $\sum_{\bm{o}}P(\bm{o}|\bm{i}) = 1$ for all $\bm{i}$.

\begin{definition}\label{def:pf}[\emph{\textbf{Process function}}] 
    A deterministic process defines a map $\bm{w}$ called the \textbf{\textit{process function}}, composed of a set of local functions $\bm{w}\coloneq (w_k)_k$ where $w_k:\mathcal{A}\rightarrow\mathcal{X}_k$,
which specifies the relation between outputs and inputs in Eq.\eqref{eq:correlation2} via $P(\bm{x}|\bm{a})=\delta_{\bm{x},\bm{w}(\bm{a})}$, where $\delta_{i,j}$ is the Kronecker delta. Similarly, deterministic local channels, also called \textbf{\textit{local interventions}}, define  functions $\bm{f}:=(f_k)_k$ where $f_k : \mathcal{X}_k \times \mathcal{I}_k \to \mathcal{A}_k \times \mathcal{O}_k$ such that $P(a_k,o_k|x_k,i_k)=\delta_{(a_k,o_k),f_k(x_k,i_k)}$. The process function scenario, assuming deterministic channels, is depicted in Fig.~\ref{fig_pf}.
\end{definition}

\subsection{Fixed point uniqueness\label{sec:characterisation_pfs}}

Let us consider the simplified scenario where $|\mathcal{I}_k|=|\mathcal{O}_k|=1$ for all parties $k$.  Therefore, Eq.~\eqref{eq:correlation2} becomes:
\begin{align}
1=\sum_{\bm{a},\bm{x}}\prod_{k=1}^{n}\delta_{a_k,f_k(x_k)}\delta_{x_k,w_k(\bm{a})} &=\sum_{\bm{x}}\prod_{k=1}^{n}\delta_{x_k, w_k(\bm{f} (\bm{x}))},\notag\\
&=\sum_{\bm{a}}\prod_{k=1}^{n}\delta_{a_k, f_k(w_k (\bm{a}))}.
\label{eq:logical_consistency_deterministic}
\end{align}
Logical consistency can hold if and only if, for any choice of local operations $\bm{f}$, the system of equations $\bm{x}=\bm{w}(\bm{f} (\bm{x}))$ has exactly one solution $\bm{x}=\bm{x^f}$ , or equivalently $\bm{a}=\bm{f}(\bm{w} (\bm{a}))$ has exactly one solution $\bm{a}=\bm{a^f}$ . This yields the following characterization:

\begin{theorem}\label{th:fixedpoint}[\emph{\textbf{Fixed point uniqueness}}] For any function $\bm{w}:\mathcal{A}\rightarrow\mathcal{X}$ that is a set of local functions $\bm{w}\coloneq (w_k)_k$ where $w_k:\mathcal{A}\rightarrow\mathcal{X}_k$, $\w$ is a process function if and only if
    \begin{align}
    &\forall \bm{f}:\mathcal{X}\rightarrow \mathcal{A},\hspace{1mm} \exists!\hspace{1mm} \bm{x^{f}}\hbox{ such that } \forall k,\ x_k^f=w_k(\bm{f}(\bm{x^f})),\label{eq:fixedpointx}\\
    &\hbox{ or equivalently, }\notag\\
    &\forall \bm{f}:\mathcal{X}\rightarrow \mathcal{A},\hspace{1mm} \exists!\hspace{1mm} \bm{a^{f}}\hbox{ such that } \forall k,\ a_k^f=f_k(w_k(\bm{a^f})).
    \label{eq:fixedpointa}
\end{align}
\begin{proof}
    Eq.~\eqref{eq:fixedpointx} is proven in Refs.~\cite{baumeler16,baumeler19}. Eq.~\eqref{eq:fixedpointa} follows from this proof and Eq.~\eqref{eq:logical_consistency_deterministic}. 
\end{proof}
\end{theorem}

This characterization gives an intuition for how process functions can give rise to structures that are incompatible with any well-defined causal order, yet distinct from ``logically inconsistent quasi-processes''~\cite{kunjwal23}. The latter lead to the \emph{grandfather antinomy}, which exhibits underdetermination\textemdash there exists a set of interventions $\bm{f}$ such that $\bm{w}\circ\bm{f}$ admits no fixed point\textemdash and to the \emph{bootstrap antinomy}, characterized by overdetermination\textemdash there exists $\bm{f}$ such that $\bm{w}\circ\bm{f}$ admits more than one fixed point. The unique fixed-point property thus appears to act as an intrinsic ``time-policing" principle that is nevertheless compatible with freedom of choice.\\

Eqs.~\eqref{eq:fixedpointx}--\eqref{eq:fixedpointa} directly imply a  \textit{non-self-signaling property}~\cite[Lemma 1]{baumeler21} (ensuring that each local component $w_k:\mathcal{A}\rightarrow\mathcal{X}_k$ of $\bm{w}$ is independent of the output $\mathcal{A}_k$ of party $k$) that guarantees that this composition is well defined, preventing an infinite recursion between $a_k=f_k(x_k)$ and $x_k=\bm{w}_k(\bm{a})$. 
In a one-party scenario, the only process function $w$ that ensures that $w\circ f$ has a unique fixed point for any choice of $f$ is a constant function: $\forall a, w(a)=x$. This means the party's input $x$ is fixed and cannot depend on its own output $a$, though it remains free to apply any operation $f$ to it.  Logical consistency thus prevents the party from signaling to its own past, which would otherwise lead to paradoxes. For instance, if $x:=w(a)=a\oplus 1$ with $x,a\in\{0,1\}$, and the local operation is defined as the identity map $\mathrm{Id}:x\rightarrow a:=x$,  this leads to the contradiction $a=a\oplus1$, i.e.~$w\circ f$ has no fixed point, which corresponds to the grandfather antinomy.  
Under free local operations, it was shown in Ref.~\cite{baumeler21} that the grandfather antinomy is equivalent to the information antinomy, i.e.~given an intervention $f$ producing the  grandfather antinomy in $w$, there exists some other intervention $f^\prime$ yielding the bootstrap antinomy. For instance,  if $x:=w(a)=a\oplus 1$ with $x,a\in\{0,1\}$, and the local operation is the bit flip map $f^\prime:x\rightarrow a:=x\oplus1$, this results in a tautology $a=a$, that is, $w\circ f^\prime$ has two fixed points, corresponding to the bootstrap antinomy.

As a consequence, any $n-$partite
process function can be expressed as a set of functions mapping the outputs of $n\backslash k$ parties into the input of the party $k$, $w_k:\mathcal{A}_{\backslash k}\rightarrow\mathcal{X}_k$, $x_k:=w_k(\bm{a}_{\backslash k})$~\cite{baumeler21}.\\

Other characterizations of process functions, based on recursivity, have been 
proposed~\cite{baumeler19,tobar}, but they do not forbid $n$-partite loops, and are thus 
necessary but not sufficient criteria. We present an alternative recursive characterization and revisit, discuss, and complete these existing approaches 
in forthcoming work.

\subsection{Causal and non-causal process functions}\label{sec:causalnon-causal}

In this section, we introduce the notion of (non-)causal correlations and propose a recursive characterization of (non-)causal process functions, which will later be used to relate non-causality to QNLWE.\\

From Section \ref{sec:characterisation_pfs}, one can see that all bipartite process functions are causal, i.e.~are always compatible with a well-defined causal order defined by the non-two-way signaling condition. This implies that bipartite process functions can only generate \emph{causal correlations}, defined as \emph{theory-independent} probability distributions that can be decomposed as a convex mixture of one-way-non-signaling probability distributions \cite{oreshkov1,araujo1,oreshkov16,branciard1},
\begin{align}
   \nonumber P(o_1,o_2|i_1,i_2) &= q P^{A_1\prec A_2}(o_1,o_2|i_1,i_2)\\
   &\qquad +(1-q)P^{A_2\prec A_1}(o_1,o_2|i_1,i_2),
\end{align}
with $q\in [0,1],$ and $A_i\prec A_j$ denoting that party $A_i$ causally precedes $A_j$, i.e.~that $A_j$ cannot signal to $A_i$ which lies in its past,  translating into the following non-signaling conditions: 
\begin{align}
  \nonumber &\forall o_1,i_1,i_2,i_2'\!:\\
  &\qquad\sum_{o_2}P^{A_1\prec A_2}(o_1,o_2|i_1,i_2)=\sum_{o_2}P^{A_1\prec A_2}(o_1,o_2|i_1,i_2'), \\
  \nonumber &\forall o_2,i_1,i_1',i_2\!: \\
  &\qquad \sum_{o_1}P^{A_2\prec A_1}(o_1,o_2|i_1,i_2)=\sum_{o_1}P^{A_2\prec A_1}(o_1,o_2|i_1',i_2).
\end{align}
The set of bipartite causal correlations is given by the convex hull of the two sets $P^{A_1 \prec A_2}$ and $P^{A_2 \prec A_1}$, forming the  \textit{causal polytope}~\cite{branciard1}, which has also been extended to the richer multipartite case~\cite{abbott16}.
Some of its facets are trivial, corresponding to the non-negativity 
constraints $P(o_1,o_2|i_1,i_2)\geq 0$, while the nontrivial facets 
define causal inequalities\textemdash linear constraints satisfied by all 
correlations compatible with a definite causal order.  Causal 
inequalities characterize the boundary of causal correlations (though not 
all such inequalities correspond to facets) \cite{oreshkov1,branciard1}. They can be expressed as 
linear combinations of the conditional probabilities $P(o_1,o_2|i_1,i_2)$, 
bounded by causal constraints 
or equivalently as causal games, where 
the inequality defines a score or success probability. A violation of a 
causal inequality demonstrates that the observed correlations are 
non-causal, i.e.~incompatible with any definite causal order. The notion of (non-)causal correlations has also been generalized to multipartite scenarios:

\begin{definition}\label{def:causalcor}
    [\emph{\textbf{Causal correlations}}]  For $n-$partite scenarios, causal correlations are defined as those that can be decomposed as \cite{oreshkov16,abbott16}
 \begin{equation}
     P(\bm{o}|\bm{i})=\sum_k q_k P_k(o_k|i_k)P_{k,i_k,o_k}(\bm{o}_{\backslash k}|\bm{i}_{\backslash k}),
     \label{eq:causalcor}
 \end{equation}
 with $q_k\geq 0$, $\sum_k q_k=1$, $P_k(o_k|i_k)$ is a single party probability distribution (hence trivially causal), $P_{k,i_k,o_k}(\bm{o}_{\backslash k}|\bm{i}_{\backslash k})$ is a $(n-1)-$partite causal correlation for each $k,i_k,o_k$. Each term in the decomposition is non-signaling with respect to the chosen party $k$ who lies in the past of the remaining ones, ensuring that they cannot signal to $k$ within that term. 
\end{definition}

 This recursive definition of causal correlations can be understood as a consequence of a causality principle, which states that a freely chosen setting cannot be correlated with any variable located in its causal past or outside its causal influence \cite{oreshkov16}.

 \medskip
 
 The special case of ``classical correlations'', i.e.\ those generated from from random variables and stochastic processes, has also been studied \cite{baumeler2}. Strikingly, even within this purely classical framework, non-causal correlations can arise for multipartite scenarios in case of $n>2$. Restricting further to deterministic dynamics (i.e.~process functions) still allows for the existence of such non-causal correlations \cite{baumeler2}.

\begin{definition}\label{def:causalpf}
    [\emph{\textbf{(Non-)Causal process function}}] A process function is causal if it generates causal correlations under any local operations, i.e.~it can never lead to the violation of a causal inequality. On the other hand,
process functions that are able to generate such non-causal correlations are themselves labeled as non-causal \cite{oreshkov1,branciard1}.
\end{definition}

 The Lugano process $\w^L$ (cf.~Eq.~\eqref{eq:lugano}) is the canonical example of a non-causal process function. It achieves the maximal violation of the causal inequality
\begin{align}
    &\frac{1}{8}\big(P(000|000) + P(100|001) + P(001|010) \notag\\
    &+ P(001|011) + P(010|100) + P(100|101) \notag \\
    &+ P(010|110) + P(000|111)\big) \leq \frac{3}{4},
    \label{eq:lugineq}
\end{align}
which is commonly presented as a causal game \cite{baumeler2}
\begin{align}
  &  \frac{1}{2}\big(P(o_1=i_3,o_2=i_1,o_3=i_2|maj(\bm{i})=0)\notag\\
  &+P(o_1=i_2\oplus 1,o_2=i_3\oplus 1,o_3=i_1\oplus  1|maj(\bm{i})=1)\big)\notag\\ &\hspace{68mm}\leq \frac{3}{4}.
\end{align}
The maximal violation is achieved when the underlying process is $\w^L$ and each party implements the local strategy $P(a_k,o_k|x_k,i_k) = \delta_{a_k,i_k}\delta_{o_k,x_k}$. Under this strategy, the parties effectively perform a swap: each feeds its external input into the process function ($a_k = i_k$) and outputs the value received from the process ($o_k = x_k$). We note moreover that
each term $P(o_1,o_2,o_3|i_1,i_2,i_3)$ in Eq.~\eqref{eq:lugineq} can be uniquely associated with an event $(x_1,x_2,x_3|a_1,a_2,a_3)$ of the Lugano process:
\begin{equation}
\label{eq:lugevent}
\begin{array}{
    c@{\,}
    l@{,\,\,}
    l@{,\,\,}
    l@{,\,\,}
    l
    @{\,}c
}
\{&
(000|000) & (100|001) & (001|010) & (001|011)
&
\\[2pt]
&
(010|100) & (100|101) & (010|110) & (000|111)
&\}.
\end{array}
\end{equation}

This scenario can thus be reinterpreted as playing the game ``guess the Lugano process.”  Indeed, Eq.~\eqref{eq:lugineq} can be rewritten as
\begin{equation}
    \tfrac{1}{8}\sum_{\bm{i}} P\big((o_1,o_2,o_3) = \w^L(i_1,i_2,i_3)|\bm{i}\big) \le \tfrac{3}{4}, 
    \label{eq:lugineq_rewritten}
\end{equation}
with a slight abuse of notation since $\w^L$ is defined on $\mathcal{A} \rightarrow \mathcal{X}$.   
Under the strategy described above, the resulting probability distribution Eq.~\eqref{eq:correlation2} wins the game with certainty, i.e.~$\tfrac{1}{8} \sum_{\bm{i}}P\big((o_1,o_2,o_3) = \w^L(i_1,i_2,i_3)|\bm{i}\big) = 1$.\\

More generally, to any process function $\bm{w}$, one can associate a causal inequality of the form
\begin{equation}
    \frac{1}{ |\mathcal{I}|}\sum_{\bm{i}}P(\bm{o}=\bm{w}(\bm{i})|\bm{i})\leq \beta^C\leq 1,
    \label{eq:causalineq}
\end{equation}
where $\beta^C$ denotes the causal bound \cite{baumeler2}, i.e.~the maximum score achievable  by causal correlations.  If  $\w$ is a causal process function, then the inequality is trivial: $\beta^C=1$, i.e.\ the optimal causal strategy reaches the algebraic maximum. If $\w$ is non-causal, $\beta^C<1$, any causal strategy is always below the algebraic maximum, and the inequality is maximally violated: the local strategy above combined with $\w$  achieves $\bm{o}=\bm{w}(\bm{i})$ with certainty.\\

From Eq.~\eqref{eq:correlation2} and Eq.~\eqref{eq:causalcor}, we can state the following characterization of causal process functions:\\ 

\begin{theorem}\label{thm:characcausalpf}
[\emph{\textbf{Characterization of causal process functions}}]  
An $n-$partite process function $\bm{w}:\mathcal{A}\to\mathcal{X}$ is \emph{causal} if and only if the following two conditions hold: 
\begin{enumerate}[(i)]
    \item There exists at least one party $k$ such that no other party can signal to it, i.e.~the corresponding input is constant,
    \begin{align}
        x_{k} = w_{k}(\bm{a}_{\backslash k}) := x_{k}^0 .
    \end{align}
    
    \item For every such party $k$ and for every $a_{k}\in\mathcal{A}_{k}$, the output-reduced process functions
    \begin{align}
        \bm{w}^{a_{k}}:\mathcal{A}_{\backslash k}\to\mathcal{X}_{\backslash k}\label{eq:output_red_pf}
    \end{align}
    
    defined as 
    \begin{align}
        \bm{w}^{a_k}(\bm{a}_{\backslash k}) := ( w_1(\bm{a}_{\backslash 1},a_k), \dots, w_{k-1}(\bm{a}_{\backslash \{k-1\}},a_k),\notag \\
    w_{k+1}(\bm{a}_{\backslash \{k+1\}},a_k), \dots, w_{n}(\bm{a}_{\backslash n},a_k) )\notag
    \end{align} 
    are themselves valid $(n-1)-$partite causal process functions.    
\end{enumerate}

\begin{proof}
For both directions, the proof proceeds by induction on the number of parties, see Appendix \ref{app:proofCausalPF} for more details.
\end{proof}
\end{theorem}

Non-causal process functions are therefore those that violate either condition $(i)$ or $(ii)$. Process functions that do not satisfy $(i)$ are said to be \textit{without global past}~\cite{kunjwal23a}: no input $x_k$ remains constant across all configurations of the other parties' outputs. 
In such cases, every party receives a signal, through the process, from at least one other party $k' \neq k$,  implying that no party lies in the global past of all others: 
\begin{align}
    \nonumber \forall k \!: \; &\exists k'\neq k, a_{k'}, a'_{k'},  \hspace{1mm} \hbox{s.t.} \hspace{1mm} \\
    & w_k(\bm{a}_{\backslash\{k,k'\}},a_{k'})\neq w_k(\bm{a}_{\backslash\{k,k'\}},a'_{k'})
\end{align}

If $\w$ is without global past and  $\forall k, |\mathcal{I}_k|=d$, Ref.(\cite[Theorem 7]{baumeler2}) shows that $\beta^C=1-\frac{1}{d^n}$ in inequality Eq.\eqref{eq:causalineq}. We generalize this result to arbitrary local input dimensions.

\begin{corollary}\label{cor:causalBound}
    For arbitrary $\mathcal{I}_k$, if $\bm{w:\mathcal{A}\to\mathcal{X}}$ is a process function non-causal without global past, the causal bound is given by
    \begin{equation}
        \beta^C = 1-\frac{1}{\prod_k |\mathcal{I}_k|} < 1.
    \end{equation}
    \begin{proof}
        See Appendix \ref{app:proofCausalBound}.
    \end{proof}
\end{corollary}

Non-causal process functions may still exhibit a partial causal structure ~\cite{abbott17}. 
 As simple examples, consider a $4-$partite Boolean process function, satisfying $(i)$ but violating $(ii)$, such that party 4 is in the global past of the others, and controls through her output  whether parties $(1,2,3)$ communicate with each other in a  fixed causal order ($a_4=0$) or via a Lugano process ($a_4=1$):
\begin{align}\label{eq:ngpf}
x_1 &:= (1-a_4)+a_4a_3(a_2\oplus1), \notag\\
x_2 &:=(1-a_4)a_1+ a_4a_1(a_3\oplus1), \notag\\
x_3 &:=(1-a_4)a_2+ a_4a_2(a_1\oplus1),\notag\\
x_4 &:=0,
\end{align}
and a $6-$partite Boolean process function without global past, similar to the previous one but where party 4 also takes part in a Lugano process, such that the parties 4, 5 and 6 are in the past of the parties 1, 2 and 3: 
\begin{align}\label{eq:ngpfb}
x_1 &:= (1-a_4)+a_4a_3(a_2\oplus1), \notag\\
x_2 &:=(1-a_4)a_1+ a_4a_1(a_3\oplus1), \notag\\
x_3 &:=(1-a_4)a_2+ a_4a_2(a_1\oplus1),\notag\\
x_4 &:=a_6(a_5\oplus 1),\notag\\
x_5 &:=a_4(a_6\oplus 1),\notag\\
x_6 &:=a_5(a_4\oplus 1).
\end{align}

Finally, we remark that for process functions\textemdash unlike general process matrices\textemdash causality and multipartite causal separability \cite{wechs} coincide. These are a priori distinct notions: causality concerns the observed correlations, requiring that they decompose into causally ordered ones, i.e., that they admit an explanation in terms of a well-defined causal order; whereas causal separability concerns the process itself, requiring that it decompose into causally ordered processes. The two differ in general, since some causally nonseparable processes (such as the quantum switch \cite{chiribella13} and its generalisations \cite{wechs1,purves21}) may nonetheless generate only causal correlations \cite{araujo1,wechs1,purves21}. For process functions, however, the two notions are one and the same. Indeed, \cite{abbott25} shows that a process function $\bm{w}:\mathcal{A}\to\mathcal{X}$ is causally separable (or ``causally definite'' \cite{abbott25}) if and only if there exists a party $k$ for which $w_{k}$ is constant and, for every operation $f_{k}:\mathcal{X}_{k}\rightarrow\mathcal{A}_{k}$, the reduced process $\w^{f_{k}}$ is itself causally separable. Our characterization of causality for process function coincides with this: since the global past $w_{k}$ is constant, $\mathcal{X}_{k}$ is a singleton, and hence every operation $f_{k}$ is constant as well.
\medskip

This concludes the discussion of fundamental notions of process functions. For the remainder of the article, we present and discuss our main results.

\section{From unambiguous product bases to process functions}\label{sec:upbpf}

First, we demonstrate that process functions can be constructed from unambiguous product bases. 
Let us consider an \textit{n-}partite Hilbert space $\mathcal{H}=\bigotimes_k\mathbb{C}^{d_k}$. Suppose that
\begin{align}
    \mathcal{S}=\{\ket{\psi^j}=\ket{\psi^j_1}\otimes...\otimes \ket{\psi^j_n}\}_{j=1}^{|S|}
\end{align} 
is a \textit{set of mutually orthogonal product vectors} in $\mathcal{H}$, where each local vector $\ket{\psi_k^{j}}\in \mathbb{C}^{d_k}$ is normalised.  

For each party $k$, the collection of local vectors is arranged into a set
\begin{equation}
    \mathcal{S}^{(k)} = \{\ket{\psi_k^{1}}, \ldots, \ket{\psi_k^{s_k}}\},
    \label{equ:s_i}
\end{equation}
with $s_k \leq |\mathcal{S}|$ distinct elements, such that any $\ket{\psi^j_k}$ with $j > s_k$ is already contained in $\mathcal{S}^{(k)}$. In turn, each local set $\mathcal{S}^{(k)}$ can be further partitioned as the disjoint union of  $|\mathcal{X}_k|$ ordered subsets,
\begin{equation}\label{eq:s_i_disjoint}
     \mathcal{S}^{(k)} = \bigsqcup_{x_k=0}^{|\mathcal{X}_k|-1} \mathcal{S}_{x_k}^{(k)},
\end{equation}
such that all vectors within a given subset are mutually orthogonal. We will interpret $x_k\in\mathcal{X}_k$ as a \emph{measurement setting} for party $k$, and the position of a vector within the corresponding subset $\mathcal{S}_{x_k}^{(k)}$ as the associated \emph{measurement outcome} $a_k\in\mathcal{A}_k$. 
Consequently, every global state $\ket{\psi^j}\in\mathcal{S}$ determines an event $(\bm{a}^j| \bm{x}^j)$, where
\begin{align}
    \bm{a}^j &=(a_1^j,\ldots,a_n^j)\in\mathcal{A}:=\bigtimes_{k=1}^n\mathcal{A}_k,\\
    \bm{x}^j &=(x_1^j,\ldots,x_n^j)\in\mathcal{X}:=\bigtimes_{k=1}^n\mathcal{X}_k,
\end{align}
represent the outcomes of a joint measurement and the corresponding measurement settings, respectively. Therefore, each party $k$ has its   local sets of outcomes $\mathcal{A}_k:=\{a_k\}$ and measurement settings $\mathcal{X}_k:=\{x_k\}$.

\medskip

As an illustrative example, consider the SHIFT basis~\eqref{eq:shift} on $\mathcal{H} = (\mathbb{C}^2)^{\otimes 3}$, for which $|\mathcal{S}| = 8$. Each party has two settings, corresponding to the computational basis ($x_k=0$) and the diagonal (quantum Fourier transform) basis ($x_k=1$). For every $k\in\{1,2,3\}$ the local set decomposes as
\begin{equation}
    S^{(k)}= S_0^{(k)} \sqcup S_1^{(k)} = \{\ket{0}, \ket{1}\} \sqcup \{\ket{+}, \ket{-}\}.
\end{equation}
The corresponding set of associated events $(\bm{a}^j| \bm{x}^j)$ is given by
\begin{equation}
\label{eq:shiftevent}
\begin{array}{
    c@{\,}
    l@{,\,\,}
    l@{,\,\,}
    l@{,\,\,}
    l
    @{\,}c
}
\{&
(000|000) & (001|100) & (010|001) & (011|001)
&
\\[2pt]
&
(100|010) & (101|100) & (110|010) & (111|000)
&\}.
\end{array}
\end{equation}

Notably, these events define the Lugano process~\eqref{eq:lugano}: $\mathcal{X}$ and $\mathcal{A}$ play respectively the roles of the parties' input and output spaces, as well as process output and input spaces. Written as $(\bm{x}^j| \bm{a}^j)$, each event Eq.~\eqref{eq:shiftevent} corresponds to a unique Lugano process event~\eqref{eq:lugevent}.

\medskip

We now impose two structural conditions on $\mathcal{S}$.

\medskip

\paragraph{\textbf{Completeness.}}
The set $\mathcal{S}$ is a \emph{complete product basis}, that is,
\[
|\mathcal{S}| = \prod_{k=1}^n d_k.
\]
In this case, each local subset $\mathcal{S}_{x_k}^{(k)}$ contains exactly $d_k = |\mathcal{A}_k|$ vectors. Completeness ensures that every joint outcome $\bm{a}^j \in \mathcal{A}$ labels exactly one global basis vector $\ket{\psi^j} \in \mathcal{S}$.

\medskip

\paragraph{\textbf{Unambiguity.}}
A product basis $\mathcal{S}$ is said to be \emph{unambiguous} if, for every party $k$, two \emph{distinct} local vectors are orthogonal if and only if they belong to the same local subset $\mathcal{S}^{(k)}_{x_k}$:
\begin{equation} \label{def:UA}    
\braket{\psi_k^{j'}}{\psi_k^{j}} = 0 \quad \Longleftrightarrow \quad x_k^{j'} = x_k^{j}.
\end{equation}
This condition coincides with \emph{\textbf{property (P)}} introduced in Refs.~\cite{augusiak11,augusiak12}. Operationally, unambiguity guarantees that each local vector $\ket{\psi_k^j}$ admits a \emph{unique} event label $(a_k^j | x_k^j)$. Indeed, suppose \textit{ad absurdum} that $\ket{\psi_k^j}$ belonged to two distinct subsets $\mathcal{S}^{(k)}_{x_k}$ and $\mathcal{S}^{(k)}_{x_k'}$ with $x_k \neq x_k'$. Then there would exist vectors $\ket{\psi_k^{j'}} \in \mathcal{S}^{(k)}_{x_k}$ and $\ket{\psi_k^{j''}} \in \mathcal{S}^{(k)}_{x_k'}$ such that
$
\braket{\psi_k^{j'}}{\psi_k^{j}} = \braket{\psi_k^{j''}}{\psi_k^{j}} = 0.$
By unambiguity, this implies $x_k^{j'} = x_k^{j} = x_k^{j''}$, and hence $x_k = x_k'$, contradicting our assumption. 
 Hence, for all $j$ and all $x_k^{j'} \neq x_k^{j}$,
\[
(a_k^j | x_k^{j'}) \not\equiv (a_k^j | x_k^j),
\]
and each local vector $|\psi_k^j\rangle$ appears in exactly one subset $\mathcal{S}^{(k)}_{x_k}$ in the partition~\eqref{eq:s_i_disjoint}. 
As a consequence, the assignment of measurement settings is unambiguous, and we may consistently identify local vectors with their event labels by writing $\ket{(a_k^j | x_k^j)} := \ket{\psi_k^j}$. We discuss weaker notions of unambiguity and their relation to the present definition in Appendix~\ref{app:weakUA}.

\medskip

Together, completeness and unambiguity ensure that each local subset $\mathcal{S}_{x_k}^{(k)}$ forms a \emph{complete local orthonormal basis} (see Appendix \ref{app:locbases}). 

Moreover, they imply that each joint outcome $\bm{a}^j$ is associated with a unique joint setting $\bm{x}^j$. Hence the basis $\mathcal{S}$ defines a well-defined unique \emph{quasi-process function}
\[
\bm{w} : \mathcal{A} \rightarrow \mathcal{X},
\qquad
\bm{w}(\bm{a}^j) = \bm{x}^j,
\]
with $\bm{w} = (w_k)_k$ and $w_k : \mathcal{A} \to \mathcal{X}_k$ (cf.~Lemma~\ref{app:lemma:funcBasis} in Appendix~\ref{app:lemmasUA}). Moreover, unambiguity implies that orthogonality of global basis vectors translates into the following exclusivity structure among the corresponding events.

\begin{lemma}\label{lem:PE}[\emph{\textbf{Pairwise Exclusivity}}]
For any two distinct events $(\bm{a}^j | \bm{x}^j)$ and $(\bm{a}^{j'} | \bm{x}^{j'})$ associated with $\mathcal{S}$, there exists at least one party $k$ such that
\begin{equation}\label{eq:PE}
    a_k^j \neq a_k^{j'} \quad \text{and} \quad x_k^j = x_k^{j'} .
\end{equation}
\begin{proof}
    Indeed, since $\mathcal{S}$ is an orthonormal product basis, orthogonality of the global states $\ket{\psi^j}$ and $\ket{\psi^{j'}}$ implies that at least one pair of local vectors $\ket{\psi_k^j}$ and $\ket{\psi_k^{j'}}$ is orthogonal. By unambiguity, this can only occur when the two vectors belong to the same local subset $\mathcal{S}^{(k)}_{x_k}$ and hence correspond to the same measurement setting but different outcomes, yielding Eq.~\eqref{eq:PE}. A detailed proof is given  in Appendix~\ref{app:lemmaPE}.
\end{proof}
    
\end{lemma}

We now show that the exclusivity structure enforced by unambiguity directly implies non-self-signaling of the associated quasi-process function.

\begin{lemma}\label{lem:nss}[\emph{\textbf{Non-self-signaling}}] The pairwise exclusivity of events arising from a complete unambiguous product basis implies that the associated quasi-process function $\bm{w}$ is non-self-signaling, i.e.~
\begin{equation} \label{eq:nss}
    \forall k,\;\forall a_k^{j'} \neq a_k^{j}\!: \quad w_k(a_k^{j'}, \bm{a}_{\backslash k}^j) = w_k(a_k^{j}, \bm{a}_{\backslash k}^j).
\end{equation}
\begin{proof}
    Consider an arbitrary party $k$ and two basis vectors $\ket{\psi^{j}}, \ket{\psi^{j'}} \in \mathcal{S}$, whose corresponding events differ only in party $k$'s outcome:
    \[
        \bm{a}^{j} = (a_{k}^{j}, \bm{a}_{\backslash k}^{j}), \qquad\bm{a}^{j'} = (a_{k}^{j'}, \bm{a}_{\backslash k}^{j}).
    \]
    By pairwise exclusivity~\eqref{eq:PE}, this implies $x_{k}^{j'} = x_{k}^{j}$, and therefore
    \[
        w_{k}({a^{j'}_{k}},{\a_{\backslash k}^{j}})=w_{k}({a^{j}_k},{\a_{\backslash k}^{j}}).
    \]
    Since $k$ is arbitrary, we conclude that $\bm{w}$ is non-self-signaling, where $\bm{w}=(w_k:\mathcal{A}_{\backslash k}\rightarrow\mathcal{X}_k)_k$.
\end{proof}
\end{lemma}

\medskip

We can now proceed to demonstrate one of our main results:

\medskip

\begin{theorem}\label{thm:unamb_to_pf}[\emph{\textbf{From unambiguous product basis to process function}}]
    Any unambiguous complete joint $n-$partite product basis determines a unique $n-$partite process function. \label{thm:btopf}
\end{theorem}

\begin{proof}
Let $\mathcal{S}$ be a complete unambiguous product basis, and let
$E=\{(\bm a^{j}\mid \bm x^{j})\}_{j}$ be its associated event list. By completeness,
every joint outcome $\bm a^j\in\mathcal A$ labels exactly one basis vector $\ket{\psi^j}$. Combined with unambiguity, this ensures that each setting
$\bm x^j=\bm w(\bm a^j)$ is well defined, i.e.\ $\w$ is total. Hence $\bm w:\mathcal A\to\mathcal X$ is a total
quasi-process function, which satisfies
pairwise exclusivity by Lemma~\ref{lem:PE} and is non-self-signaling by Lemma~\ref{lem:nss}.

To probe the logical consistency of $\bm w$, we allow the parties to intervene: each party $k$ freely applies a local map $f_k:\mathcal X_k\to\mathcal A_k$, turning the setting it receives into an outcome. Collecting these into $\bm f=(f_k)_k$, this collection admits a fixed point if there exist $\bm a$ such that $\bm a=\bm f(\bm w(\bm a))$ (Eq.~\eqref{eq:fixedpointa}). Since $\bm w$ is total, this equation is well posed for every $\bm f$, and any fixed point is itself an outcome labelling a basis vector, i.e.~an event of $E$. Assume \emph{ad absurdum} that, for some intervention $\bm f$, two distinct basis vectors $\ket{\psi^j},\ket{\psi^{j'}}\in\S$ have labels $\bm a^j\neq\bm a^{j'}$ that are
both fixed points, i.e.~$a_k^{j}=f_k(w_k(\bm a^{j}))$ and $a_k^{j'}=f_k(w_k(\bm a^{j'}))$ for all $k$. By
pairwise exclusivity, there is a party $k$ with $a_k^j\neq a_k^{j'}$ and $w_k(\bm a^j)=w_k(\bm a^{j'})$, but the latter implies
\begin{equation}
  a_k^{j}=f_k(w_k(\bm a^{j}))=f_k(w_k(\bm a^{j'}))=a_k^{j'},
\end{equation}
a contradiction. Hence for every intervention $\bm f$ there exists \emph{at most one}
fixed point $\bm a=\bm f(\bm w(\bm a))$. The quasi-process function $\w$ does not suffer from the bootstrap antinomy (it is a ``pseudo-process function'' \cite{baumeler21}).

By (Ref.~\cite{baumeler21}, Theorem~1)\footnote{The terminology of Baumeler-Tselentis \cite{baumeler21} is the following: their ‘induced function’ corresponds to our quasi-process function; their ‘process function’ corresponds to a quasi-process function
that admits at least one fixed point for every choice of intervention; their ‘pseudo-process function’ corresponds to a quasi-process function that admits at most one fixed point for
every choice of intervention.  Their key result, relevant here, is that a quasi-process function is a ‘process function’ if and only if it is a ‘pseudo-process function’; as a corollary, they establish that a quasi-process function admits a grandfather antinomy if and only if it admits an bootstrap antinomy.}, any total quasi-process function $\w$ that does not suffer from the bootstrap antinomy has a unique fixed point, i.e.\ if $\bm f\circ\bm w$ (equivalently $\bm w\circ\bm f$, by Theorem \ref{th:fixedpoint}) admits at most one fixed point for every
intervention $\bm f$, then it admits  a unique fixed point for every $\bm f$. Therefore, $\bm w$ is a process function.

\end{proof}

\medskip

In Ref.~\cite{kunjwal23a}, a quantum circuit scheme was proposed to realize, via a SHIFT 
measurement, the classical channel underlying the Lugano process\textemdash namely, the map taking 
three inputs $(a_1, a_2, a_3)$ to the three outputs defined in Eq.~\eqref{eq:lugano}. When this 
channel is combined with local feedback loops from outputs to inputs, placing $(x_1, x_2, x_3)$ 
in the past of $(a_1, a_2, a_3)$, it reproduces the (classical) P-CTC representation of the 
Lugano process~\cite{araujo3, guerin, Wechs23} (see also Fig.~\ref{fig_pf}).

\begin{figure}[ht]
	\begin{center}
	\includegraphics[width=0.9\columnwidth]{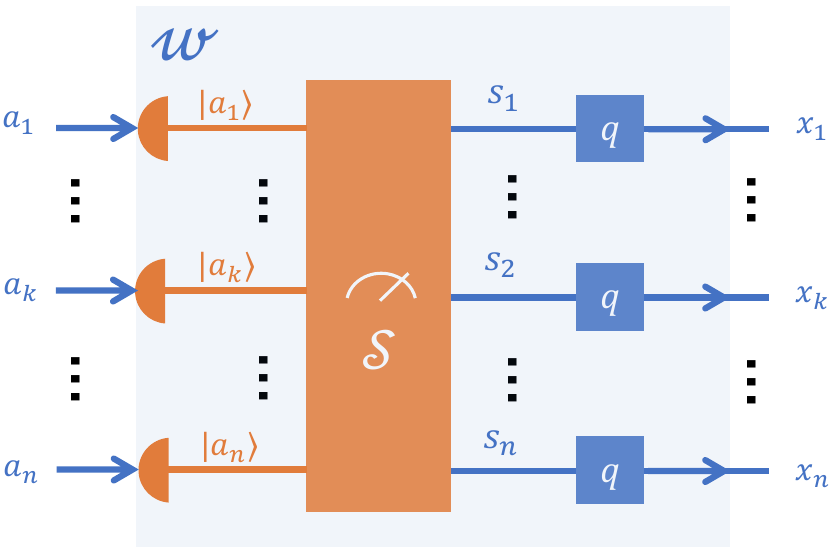}
	\end{center}
	\caption{Circuit implementing the classical channel   $\text{\calligra w}$ underlying the process function $w$ from the projective measurement on $\mathcal{S}$. For an unambiguous QNLWE basis $\mathcal{S}$, this circuit can be interpreted as a simulation of the non-causal process function, which would be genuinely implemented  if each $x_k$ were respectively in the local pasts of $a_k$ (cf.~Fig.\ref{fig_pf}).} 
\label{fig:pfchannel}
\end{figure}

Our result enables a generalized protocol that implements the classical channel associated with 
any (non-causal) process function derived from an unambiguous (QNLWE) product basis~$\mathcal S$ 
(cf.~Fig.~\ref{fig:pfchannel}). Each party encodes its classical input~$a_k$ in the 
computational basis by preparing the state~$\ket{a_k}$, which is sent to a central device 
performing a joint measurement in the basis~$\mathcal S$. Since $\mathcal S$ is a product basis, the outcome factorizes across parties, and the device returns to each party~$k$ a classical 
label~$s_k$ identifying its local post-measurement state $\ket{\psi_k^{s_k}}$ within the local set $\mathcal S^{(k)}$ of Eq.~\eqref{equ:s_i}. The process channel is then recovered by post-processing each label through a deterministic 
function $q$ defined by $q(s_k)=x_k$, where $x_k$ is the setting labeling the subset 
$\mathcal S^{(k)}_{x_k}\ni\ket{\psi_k^{s_k}}$. This function is well defined precisely because 
$\mathcal S$ is unambiguous: unambiguity makes pairwise exclusivity an equivalence relation, so 
the subsets $\{\mathcal S^{(k)}_{x_k}\}_{x_k}$ form a partition of $\mathcal S^{(k)}$ and each 
local vector belongs to exactly one of them. Hence $q$ assigns to every outcome $s_k$ a unique 
setting $x_k$. Combined with local feedback loops placing $\x$ in the past of $\a$, the composed map $\bm a\mapsto\bm s\mapsto\bm x$ reproduces the process function~$\w$.\\

Finally, we note that the unambiguity condition (Eq.\eqref{def:UA}) is automatically satisfied in the qubit case, i.e.\ whenever $d_k=2$ for all $k$. This has a striking consequence:
\begin{corollary}
    Every complete product multi-qubit basis induces a unique Boolean process function.
\end{corollary} 
It thereby establishes the exact converse of the theorem of Ref.\cite{kunjwal23a}. In the following, we will show that any valid process function, with arbitrary local (finite) dimensions and any number of parties, can also be encoded into an unambiguous product bases, thus demonstrating their equivalence.

\medskip

\section{From process functions to unambiguous product bases}\label{sec:pfupb} 
It was shown in Ref.~\cite{kunjwal23a} that a multi-qubit QNLWE basis can be constructed for every Boolean process function without a global past.  
This construction yields the SHIFT basis (Eq.~\eqref{eq:shift}), derived from the Lugano process (Eq.~\eqref{eq:lugano}).  
In this section, we generalize this result to arbitrary process functions, irrespective of the dimensions of the input and output spaces~$\mathcal{X}$ and~$\mathcal{A}$.\\

Consider a $n-$partite scenario, similar to Section~\ref{sec:pf}, in which operationally isolated parties communicate exclusively through a process function 
$\bm{w}\!:\!\mathcal{A}\!\rightarrow\!\mathcal{X}$, defined by a set of local functions 
$\bm{w} := (w_k)_k$.  
Each party $k \in \{1, \ldots, n\}$ receives a classical input $x_k \in \mathcal{X}_k$ from the process function and, in turn, provides a classical output 
$a_k \in \mathcal{A}_k$ back to it.  
In addition, each party produces an external classical outcome $o_k \in \mathcal{O}_k$, while instead of a local classical input register $\mathcal{I}_k$, it now receives an external quantum input system evolving in a Hilbert space $\HS^{I_k}$. 
Each party performs a single local quantum operation 
\begin{align}
   M_k: \mathcal{X}_k \times \mathcal{L}(\HS^{I_k}) \rightarrow \mathcal{A}_k \times \mathcal{O}_k 
   \label{eq:localpovm}
\end{align}
where $\mathcal{L}(\HS^{X})$ denotes the space of linear operators on $\HS^{X}$ and we write $\HS^{XY} := \HS^{X} \otimes \HS^{Y}$ for brevity.  
Operationally, $M_k$ is represented by a positive-operator-valued measure (POVM)
$M_k := (M_{a_k,o_k|x_k}^{I_k})_{a_k,o_k}$, where each element 
$M_{a_k,o_k|x_k}^{I_k} \in \mathcal{L}(\HS^{I_k})$ satisfies
$M_{a_k,o_k|x_k}^{I_k} \ge 0$ and 
$\sum_{a_k,o_k} M_{a_k,o_k|x_k}^{I_k} = \openone^{I_k}$ for all~$x_k$. \\

Given a separable state 
$\bm{\rho} := \bigotimes_k \rho_k$, 
with each $\rho_k \in \mathcal{L}(\HS^{I_k})$, in analogy with Eq.\eqref{eq:correlation2},
the probability distribution generated in a \emph{local operations with process function} (LOPF)  scenario is
\begin{align}
P(\bm{o}|\bm{\rho})
=\sum_{\bm{a},\bm{x}} 
\bigotimes_{k=1}^{n} 
\Tr\!\left[
M_{a_k,o_k|x_k}^{I_k}\,
\rho_k
\right]
\delta_{\bm{x},\bm{w}(\bm{a})},
\label{eq:correlation}
\end{align}
with $P(\bm{o}|\bm{\rho}) \ge 0$ and 
$\sum_{\bm{o}} P(\bm{o}|\bm{\rho}) = 1$
under arbitrary local strategies defined by
\begin{align}
P(a_k,o_k|x_k,\rho_k)
= \Tr\!\left[
M_{a_k,o_k|x_k}^{I_k}\,
\rho_k
\right].
\end{align}

We assume that the outcome register $\mathcal{O}_k$ is isomorphic to $\mathcal{A}_k$ and that, for all $k$ and $x_k$, one has $o_k = a_k$.  
This assumption implies that each local measurement produces a unique classical outcome that is both sent to the process and recorded externally.  Using the simplified notation 
$M_{a_k|x_k}^{I_k}:=\delta_{o_k,a_k}M_{a_k,o_k|x_k}^{I_k}$, Eq.\eqref{eq:correlation} becomes

\begin{align}
P(\bm{a}|\bm{\rho})
= \Tr\!\left[(E_{\bm{a}}^{\bm{I}}) \bm{\rho}\right] 
\end{align}
with
\begin{align}
E_{\bm{a}}^{\bm{I}} 
= \bigotimes_{k} 
\,M_{a_k|w_k(\bm{a}_{\backslash k})}^{I_k}.
\label{eq:dpvm}
\end{align}

\begin{figure}[ht!]
	\begin{center}
	\includegraphics[width=0.9\columnwidth]{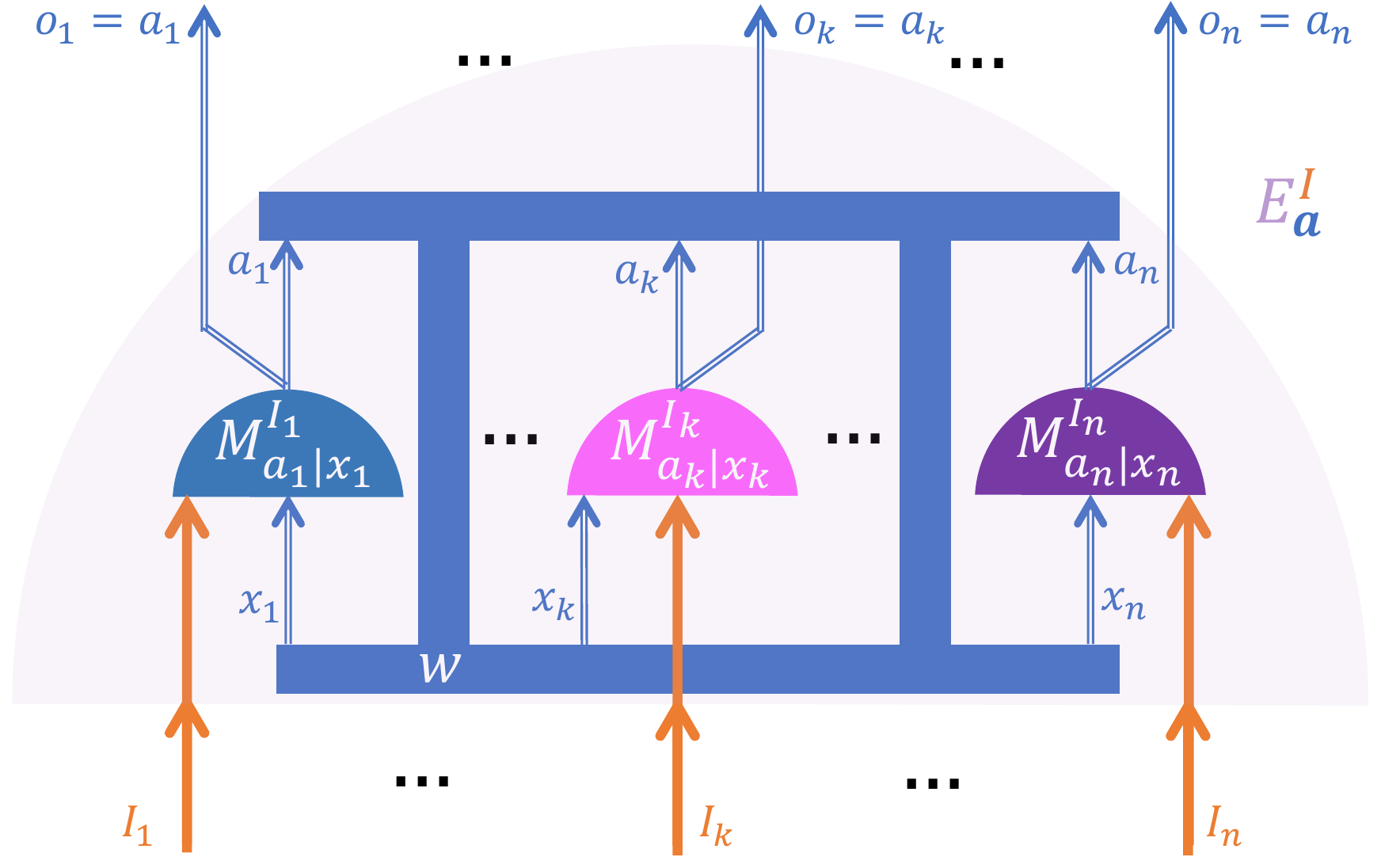}
	\end{center}
	\caption{The LOPF scenario: $n$ parties perform local (projective) operations $(M_{a_k|x_k}^{I_k})_{a_k}$ for each $k$, on separated quantum systems $\HS^{I_k}$ (orange wires). The process function $w$ (blue comb) maps the parties' outputs $\bm{a}$ into their respective inputs $\bm{x}$. This set-up generates an effective measurement $(E_{\bm{a}}^{\bm{I}})_{\bm{a}}$ (cf.~Eq.~\eqref{eq:dpvm}). In Theorem~\ref{thm:pftob}, we demonstrate that when for each $k$, $M_{a_k|x_k}^{I_k} := \ketbra{(a_k|x_k)}{(a_k|x_k)}^{I_k}$, this measurement is precisely a projection on an unambiguous complete product basis $\S$.  } 
	\label{fig:lopf}
\end{figure}

In Appendix~\ref{app:proofth1}, we show that the set of operators $(E_{\bm{a}}^{\bm{I}})_{\bm{a}}$, resulting from implementing these local quantum operations within the process function,
constitutes an effective \emph{distributed projective-valued measurement}~\cite{supic17,hoban18,dourdent21,dourdent24}, 
which we refer to as a \textit{LOPF measurement} (Fig.~\ref{fig:lopf}). \\

Finally, consider that  each party performs a fixed projective measurement in a local basis labeled by~$x_k$, with projectors 
\begin{align}
M_{a_k|x_k}^{I_k} := \ketbra{(a_k|x_k)}{(a_k|x_k)}^{I_k},
\label{eq:localpvmelmt}
\end{align}
with $M_{a_k|x_k}^{I_k} \ge 0$, $\sum_{a_k} M_{a_k|x_k}^{I_k} = \openone^{I_k}$, and $(M_{a_k|x_k}^{I_k})^2 = M_{a_k|x_k}^{I_k}$.  
Here, $x_k \in \mathcal{X}_k$ and $a_k \in \mathcal{A}_k$ denote, respectively, the measurement setting provided by the process function and the corresponding outcome of party~$k$.\\

We now show that this LOPF scenario naturally induces a joint separable projective measurement whose associated product basis is both complete and unambiguous.

\begin{theorem}\label{thm:pftob}[\emph{\textbf{From process function to unambiguous product basis}}]
Let $\bm{w} := (w_k)_k$ be an $n-$partite process function.  
Define the set $\mathcal{S}$ of product vectors generated by local applications of controlled unitaries 
$U_k^{(w_k(\bm{a}_{\backslash k}))} \equiv U_k^{x_k}$ 
on the computational basis $\{ \ket{\bm{a}} \in \mathcal{A} \}_{\bm{a}}$ as\footnote{To simplify the notation, we omit the index 
$j$ used previously in Section \ref{sec:upbpf} and write $\a:=\a^j$, $a_k^{(')}:=a_k^{j^{(')}}$, $x_k^{(')}:=x_k^{j^{(')}}$.}
\begin{align}
\mathcal{S} 
= 
\big\{
\big(\bigotimes_{k} U_k^{(w_k(\bm{a}_{\backslash k}))}\big)
\ket{\bm{a}} 
~\big|~ 
\bm{a} \in \mathcal{A}
\big\}.
\label{eq:pftoupb}
\end{align}
If, for all $a_k' \neq a_k$, the condition
\begin{align}\bra{a_k'} (U_k^{x_k'})^\dagger U_k^{x_k} \ket{a_k} = 0 \quad \text{iff}\quad x'_k=x_k\label{eq:unambunit}\end{align}
%
holds, then $\mathcal{S}$ forms a complete and unambiguous product basis.
\end{theorem}


\textit{Proof--}
Each vector in $\mathcal{S}$ uniquely corresponds to an element $E_{\a}^{\bm{I}}=\bigotimes\limits_k\ketbra{(a_k|w_k(\a_{\backslash k}))}{(a_k|w_k(\a_{\backslash k}))}^{I_k}$. with $\ket{(a_k|x_k)}^{I_k} = U_k^{(w_k(\bm{a}_{\backslash k}))}\ket{a_k}^{I_k}$. We show in Appendix \ref{app:proofth1} that the set of operators $(E_{\a}^{\bm{I}})_{\a}$  is an effective distributed projective-valued measurement. 
Thus, $\mathcal{S}$ is a \textit{complete} orthonormal basis.  

Unambiguity (Eq.~\eqref{def:UA}) requires each local vector 
$\ket{(a_k|w_k(\bm{a}_{\backslash k})}^{I_k}$ 
to be associated with a unique local basis~$x_k$, and the local bases to be disjoint.  
Operationally, this means that for each party, the sets of local measurement operators 
$(M_{a_k|x_k}^{I_k})_{a_k}$ must be disjoint, ensuring that the same state 
$\ket{(a_k|x_k)}^{I_k}$ cannot appear in two different measurement contexts, and that orthogonality only occurs within a given local basis.  Eq.~\eqref{eq:unambunit}  leads directly to this constraint on the choice of local unitaries.  \hfill$\square$\\

While this requirement is automatically satisfied in the Boolean case 
$\bm{x}, \bm{a} \in \{0,1\}^n$\textemdash for instance, by taking 
$U_k^{x_k = 0} = \openone$ and $U_k^{x_k = 1} = H$ (the Hadamard operator)\textemdash it does not hold in general for higher dimensions $d_k > 3$. In such cases, Eq.~\eqref{eq:unambunit} can still be satisfied by fixing $U_k^{0}=\openone$ and choosing the remaining unitaries $\{U_k^{x_k}\}_{x_k\neq 0}$ so that no unitary leaves any computational basis vector invariant, and such that no pair of distinct unitaries shares a common basis direction. In other words, each $U_k^{x_k}$ maps every computational basis state to a superposition with full support, i.e.~with nonzero amplitude on every computational basis
state. This is exactly what enforces unambiguity: since
$\braket{a_k'}{U_k^{x_k}|a_k}\neq 0$ for all $a_k,a_k'$, no vector of setting $x_k$ is orthogonal to any computational ($x_k=0$) basis vector, so orthogonality can never occur across two local sets, only within a set, between distinct outcomes of the same setting. The canonical full-support unitary is the quantum Fourier transform, whose matrix elements $\tfrac{1}{\sqrt{d_k}}\,e^{2\pi i\,a_k a_k'/d_k}$ all share the modulus
$1/\sqrt{d_k}$ (reducing to the Hadamard for $d_k=2$). The same property must hold pairwise for any two settings $x_k\neq x_k'$: the transition matrix
$\bra{a_k'}U_k^{x_k'\dagger}U_k^{x_k}\ket{a_k}$ must have no vanishing entry, so that vectors from different settings always overlap and are never orthogonal, as required by unambiguity.

\medskip

Although a given unambiguous product basis defines a unique process function (Theorem~\ref{thm:btopf}), distinct unambiguous product bases may correspond to the same process function, or equivalently, a single process function may admit multiple unambiguous product-basis representations. This redundancy originates from the freedom in choosing local unitaries, provided that the unambiguity condition is maintained (Theorem~\ref{thm:pftob})\footnote{In contrast, certain ambiguous product bases may also correspond to valid process functions; however, they simultaneously admit inconsistent event labeling. Other ambiguous bases (such as the domino basis \cite{bennett_99}) can never represent a valid process function  (see Appendix~\ref{app:ambi}).
}. 

\section{Non-causality and QNLWE are equivalent}\label{sec:ncqnlwe}

Theorems~\ref{thm:btopf} and~\ref{thm:pftob} together establish a one-to-one correspondence between process functions and equivalence classes of unambiguous product bases. In the special case of non-causal process functions and QNLWE bases, this immediately gives the following corollary:

\begin{corollary}\label{cor:nlwenoncauslity}[\emph{\textbf{Non-causality - QNLWE equivalence}}]
Any non-causal process function is in one-to-one correspondence with an equivalence class of unambiguous QNLWE product bases.
    \begin{proof}
   This follows directly from the above theorems and from the definition of a QNLWE basis as one that is not measurable by local operations with (causal) classical communication.
\end{proof}
\end{corollary}

These results provide a systematic construction of a process function from any unambiguous complete product basis, and conversely.

\subsection{Construction of non-causal process functions}

Given an unambiguous product basis $\mathcal{S}$, one first partitions each local set $\mathcal{S}^{(k)}$ into local bases. Then, for each party $k$, one determines the corresponding function $w_k$ by identifying for which values of $\a_{\backslash k}$ the equations $w_k(\a_{\backslash k}) = x_k$ holds, for all possible values of $x_k$.   
As an illustration, consider the three-qutrit QNLWE basis  (based on \cite{niset06}):
\begin{equation}
\label{eq:qutrit_lugano_likemain}
\begin{array}{
    l@{\;}
    l@{}l@{}l@{,\quad}
    l@{}l@{}l@{,\quad}
    l@{}l@{}l
    @{\;}l
}
\{&
\ket{0} & \ket{0} & \ket{0}
& \ket{\alpha_0} & \ket{0} & \ket{1}
& \ket{\beta_0} & \ket{0} & \ket{2}
&
\\[2pt]
&
\ket{0} & \ket{1} & \ket{\alpha_0}
& \ket{0} & \ket{1} & \ket{\alpha_1}
& \ket{0} & \ket{1} & \ket{\alpha_2}
&
\\[2pt]
&
\ket{0} & \ket{2} & \ket{\beta_0}
& \ket{0} & \ket{2} & \ket{\beta_1}
& \ket{0} & \ket{2} & \ket{\beta_2}
&
\\[2pt]
&
\ket{1} & \ket{\alpha_0} & \ket{0}
& \ket{\alpha_1} & \ket{0} & \ket{1}
& \ket{\beta_1} & \ket{0} & \ket{2}
&
\\[2pt]
&
\ket{1} & \ket{\alpha_1} & \ket{0}
& \ket{1} & \ket{1} & \ket{1}
& \ket{1} & \ket{1} & \ket{2}
&
\\[2pt]
&
\ket{1} & \ket{\alpha_2} & \ket{0}
& \ket{1} & \ket{2} & \ket{1}
& \ket{1} & \ket{2} & \ket{2}
&
\\[2pt]
&
\ket{2} & \ket{\beta_0} & \ket{0}
& \ket{\alpha_2} & \ket{0} & \ket{1}
& \ket{\beta_2} & \ket{0} & \ket{2}
&
\\[2pt]
&
\ket{2} & \ket{\beta_1} & \ket{0}
& \ket{2} & \ket{1} & \ket{1}
& \ket{2} & \ket{1} & \ket{2}
&
\\[2pt]
&
\ket{2} & \ket{\beta_2} & \ket{0}
& \ket{2} & \ket{2} & \ket{1}
& \ket{2} & \ket{2} & \ket{2}
&\}.
\end{array}
\end{equation}

Here, $\{\ket{\alpha_i}\}_i$ and $\{\ket{\beta_i}\}_i$ denote two distinct qutrit bases such that 
$\forall i,j \in \{0,1,2\}$, $\braket{\alpha_i}{j} \neq 0$, $\braket{\beta_i}{j} \neq 0$, $\braket{\alpha_i}{\beta_j} \neq 0$, i.e.~$\S$ satisfies unambiguity~\eqref{def:UA}. Moreover, it is also complete, $|\S|=27$.  
For every party~$k$, the local set is 
\[
\mathcal{S}^{(k)} = 
\{\ket{0}, \ket{1}, \ket{2}\}
\sqcup
\{\ket{\alpha_0}, \ket{\alpha_1}, \ket{\alpha_2}\}
\sqcup
\{\ket{\beta_0}, \ket{\beta_1}, \ket{\beta_2}\}.
\]
Because the basis is symmetric under party permutation, it suffices to analyze one representative party, say $k=1$.  

For party~1, $\ket{\psi_1} \in \mathcal{S}^{(1)}_{x_1 = 1} := \{\ket{\alpha_i}\}_i$, i.e.~$x_1 = 1$, whenever $(a_2,a_3) = (0,1)$;  
and $\ket{\psi_1} \in \mathcal{S}^{(1)}_{x_1 = 2} := \{\ket{\beta_i}\}_i$, i.e.~$x_1 = 2$, whenever $(a_2,a_3) = (0,2)$.  
In all other cases, $x_1 = 0$.  
Hence, the corresponding component of the process function is
\[
x_1 := \delta_{a_2,0}\big(\delta_{a_3,1} + 2\,\delta_{a_3,2}\big).
\]
By applying the same reasoning to the other two parties, we obtain the complete process function (without global past) as
\begin{align}
x_1 &:= \delta_{a_2,0}\big(\delta_{a_3,1} + 2\,\delta_{a_3,2}\big), \notag\\
x_2 &:= \delta_{a_3,0}\big(\delta_{a_1,1} + 2\,\delta_{a_1,2}\big), \notag\\
x_3 &:= \delta_{a_1,0}\big(\delta_{a_2,1} + 2\,\delta_{a_2,2}\big).
\label{eq:lugano3}
\end{align}

Although structurally similar to the Lugano process, 
this tripartite process function is, to our knowledge, 
the first explicit example of a non-causal process function 
in a multi-trit scenario.

\subsection{Construction of QNLWE unambiguous basis}

Conversely, given a process function $\bm{w}$, an unambiguous product basis can be constructed via Eq.~\eqref{eq:pftoupb} by choosing unitaries satisfying Eq.~\eqref{eq:unambunit}.\medskip

Let us note that if the process function is causal, the LOPF framework reduces to the deterministic subclass of one-round local operations and classical communication (LOCC$_1$) \cite{dutra26}, i.e.~the associated basis $\S$ is measurable by LOCC$_1$.  If a process function $\w$ is non-causal, the joint separable measurement $(E_{\bm{a}}^{\bm{I}})_{\bm{a}}$ belongs to the LOPF class but not to the single-round LOCC$_1$ class. In the special case when a process function has no global past, no party can choose a measurement basis independently of the others’ measurements. In particular, no party can initiate the communication. Therefore for all parties~$k$, $|\mathcal{X}_k|\neq 1$ and
    \begin{align}
        \mathcal{S}^{(k)} = \bigsqcup_{x_k = 0}^{(|\mathcal{X}_k| - 1)} \mathcal{S}^{(k)}_{x_k}, \label{eq:qnlwe}
    \end{align}
in which case no component of the associated process function remains constant.  

\medskip
  
As illustrations, using Eq.~\eqref{eq:pftoupb}, one can recover the SHIFT basis from the Lugano process, and the QNLWE basis of Eq.~\eqref{eq:qutrit_lugano_likemain} from the non-causal process function in Eq.~\eqref{eq:lugano3}.  
Consider for instance another example, the following four-partite non-causal process function:
\begin{align}
x_1 &:= \delta_{a_2,a_3,a_4}, \notag\\
x_2 &:= (1 - \delta_{a_1,0})\,a_4\,(a_3 \oplus 1), \notag\\
x_3 &:= (1 - \delta_{a_1,1})\,a_2\,(a_4 \oplus 1), \notag\\
x_4 &:= (1 - \delta_{a_1,2})\,a_3\,(a_2 \oplus 1),
\end{align}
where $x_k \in \{0,1\}$ for all~$k$, $a_{k \neq 1} \in \{0,1\}$, and $a_1 \in \{0,1,2\}$.  
Choosing the local unitaries 
\[
U_k^{x_k = 0} = \openone, \qquad 
U_{k \neq 1}^{x_k = 1} = H, \qquad 
U_{4}^{x_1 = 1} = H_3,
\]
with the qutrit Hadamard (quantum Fourier transform) operator $H_3$ acting as
\[
H_3 \ket{a_1^j} = \ket{\alpha_j}, \quad 
\text{where } 
\begin{cases}
\ket{\alpha_0} = \tfrac{1}{\sqrt{3}}(\ket{0} + \ket{1} + \ket{2}),\\[2pt]
\ket{\alpha_1} = \tfrac{1}{\sqrt{3}}(\ket{0} + \omega\ket{1} + \omega^2\ket{2}),\\[2pt]
\ket{\alpha_2} = \tfrac{1}{\sqrt{3}}(\ket{0} + \omega^2\ket{1} + \omega\ket{2}),
\end{cases}
\]
with $\omega = e^{2\pi i/3}$;
application of Eq.~\eqref{eq:pftoupb} yields the QNLWE basis
\begin{equation}
\begin{array}{
    c@{\;}
    l@{,\quad}
    l@{,\quad}
    l@{,\quad}
    l
    @{\;}c
}
\{&
\ket{\alpha_0 000} & \ket{0001} & \ket{001{+}} & \ket{001{-}}
&
\\[2pt]
&
\ket{01{+}0} & \ket{0101} & \ket{01{-}0} & \ket{\alpha_0 111}
&
\\[2pt]
&
\ket{\alpha_1 000} & \ket{1{+}01} & \ket{101{+}} & \ket{101{-}}
&
\\[2pt]
&
\ket{1100} & \ket{1{-}01} & \ket{1110} & \ket{\alpha_0 111}
&
\\[2pt]
&
\ket{\alpha_2 000} & \ket{2{+}01} & \ket{2010} & \ket{2011}
&
\\[2pt]
&
\ket{21{+}0} & \ket{2{-}01} & \ket{21{-}0} & \ket{\alpha_2 111}
&\}.
\end{array}
\end{equation}

One can also construct a QNLWE basis from a non-causal process function \textit{with} global past, such as the Boolean process function
\begin{align}
x_1 &:= 0, \notag\\
x_2 &:= (1 - a_1)a_4(a_3 \oplus 1) + a_1 a_3(a_4 \oplus 1), \notag\\
x_3 &:= (1 - a_1)a_2(a_4 \oplus 1) + a_1 a_4(a_2 \oplus 1), \notag\\
x_4 &:= (1 - a_1)a_3(a_2 \oplus 1) + a_1 a_2(a_3 \oplus 1),
\end{align}
in which the outcome of party~1, $a_1$, classically controls two distinct Lugano processes.  
Applying Eq.~\eqref{eq:pftoupb} and choosing
\[
U_k^{x_k = 0} = \openone, \qquad U_k^{x_k = 1} = H,
\]
leads to the QNLWE basis
\begin{equation}
\begin{array}{
    c@{\;}
    l@{,\quad}
    l@{,\quad}
    l@{,\quad}
    l
    @{\;}c
}
\{&
\ket{0000} & \ket{0{+}01} & \ket{001{+}} & \ket{001{-}}
&
\\[2pt]
&
\ket{01{+}0} & \ket{0{-}01} & \ket{01{-}0} & \ket{0111}
&
\\[2pt]
&
\ket{1000} & \ket{10{+}1} & \ket{1{+}10} & \ket{10{-}1}
&
\\[2pt]
&
\ket{110{+}} & \ket{110{-}} & \ket{1{-}10} & \ket{1111}
&\}.
\end{array}
\end{equation}

\section{Further applications}

Beyond the construction of new non-causal process functions and unambiguous QNLWE bases, our correspondence between process functions and equivalence classes of unambiguous product bases (Theorems \ref{thm:btopf} and \ref{thm:pftob}) yields various unexpected results.

\subsection{On bipartite unambiguous product bases}\label{sec:noqnlwe}

The established correspondence has direct implications for the study of QNLWE. In particular:
\begin{corollary}\label{cor:noqnlwe}
No bipartite unambiguous QNLWE basis can exist.
\begin{proof}
    All bipartite process functions are necessarily causal. By the correspondence between 
    process functions and unambiguous product bases established in Theorems~\ref{thm:btopf} 
    and~\ref{thm:pftob}, every bipartite unambiguous product basis is therefore perfectly 
    distinguishable by (single-round) LOCC.
\end{proof}
\end{corollary}
This corollary generalizes the well-known fact that no two-qubit product basis exhibits 
QNLWE~\cite{Walgate02} to arbitrary local dimensions.

\subsection{On unextendible product bases.}\label{sec:upb}

Remarkably, our construction (Theorem~\ref{thm:btopf}) is formally identical to the one 
introduced by Augusiak, Acín, \textit{et al.} in their study of Bell inequalities without 
violation~\cite{augusiak11, augusiak12}, where the unambiguity condition is introduced as ``property~(P)''. The connection runs deeper than a shared construction. To every process function one can associate, on the one hand, a causal inequality (Eq.~\eqref{eq:causalineq}) 
and, on the other, a Bell inequality that admits no violation---a \emph{non-signaling 
inequality} of the form $\sum_{\bm i}P(\bm i\,|\,\bm o=\w(\bm i))\leq 1$. The latter are 
trivial instances of \emph{local orthogonality} inequalities~\cite{fritz13}, which stem from 
the multipartite principle that events sharing distinct outcomes of the same local 
measurement are mutually exclusive, so that their probabilities sum to at most one.

This correspondence lets us translate results between the two scenarios (causal and non-signaling). In particular, the nontrivial local orthogonality inequalities derived from unambiguous UPBs~\cite{augusiak11, 
augusiak12}\textemdash those for which the local-orthogonality and non-signaling bounds differ\textemdash turn 
out to have a clear causal interpretation, as stated below.

\begin{corollary}\label{cor:upb}
Every nontrivial local orthogonality inequality derived from an unambiguous UPB~\cite{augusiak11, 
augusiak12}\textemdash i.e.\ one whose local-orthogonality and non-signaling bounds diverge\textemdash corresponds 
to a causal inequality whose maximal violation is attained only by an invalid quasi-process  function; no valid process function achieves it.
\end{corollary}
\begin{proof}
A UPB is, by definition, an incomplete product basis, i.e.\ $|\mathcal S|<\prod_k d_k$. 
Hence the map of Theorem~\ref{thm:btopf} assigns to it only a quasi-process function $\w$, 
which fails to be a valid process function. The associated local orthogonality inequality 
$\sum_{\bm i}P(\bm i\,|\,\bm o=\w(\bm i))\leq\beta^{\mathrm{class}}<1$ is nontrivial precisely 
because its classical bound $\beta^{\mathrm{class}}$ is strictly below the non-signaling bound 
$1$ ~\cite{augusiak11, 
augusiak12} ;
its maximal violation therefore requires realizing $\w$, which no valid process function 
can achieve.
\end{proof}

For example, the tripartite SHIFT UPB mentioned in the Introduction is associated with a  quasi-process describing a global causal loop, and hence cannot be realized by any valid 
process function.

\section{Discussion}\label{sec:disc}

In this work, we have established an equivalence between \emph{process functions}, namely the class of one-round deterministic classical communication that generate valid probability distributions under arbitrary local operations, and \emph{unambiguous complete product bases}, namely product bases in which every local state belongs to a unique local basis. While the latter notion was originally introduced in quantum information theory, where local indistinguishability of product states has found applications in tasks such as data hiding~\cite{divincenzo02,eggeling02}, the former arose in a very different context: as a paradox-free, information-theoretic model of  CTCs~\cite{baumeler21,tobar} , and more generally as the class of deterministic classical processes with indefinite causal order that remain logically consistent. In particular, process functions are precisely those deterministic classical communication structures that relaxed the idea of well-defined causal structures while still being in accordance with the \emph{no new physics} and \emph{freedom of choice} principles.

\medskip

By establishing a direct correspondence between these two frameworks, our results open new avenues in both directions, enabling systematic constructions and analyses that were previously unavailable, as well as a new interpretation of these phenomena.

\medskip

On the one hand, our second construction (Theorem~\ref{thm:pftob}) establishes a direct bridge between unambiguous QNLWE bases and process functions.  
In particular, it can offer new insights on QNLWE from the study of process functions (e.g.\ Corollary \ref{cor:noqnlwe}). 
More generally, this construction provides a straightforward method for certifying whether a given quasi-process function is valid, simply by testing whether it can be encoded in an unambiguous product basis via our formalism.

\medskip
On the other hand, our first construction (Theorem~\ref{thm:btopf}) offers new insight into process functions through the lens of unambiguous product bases, allowing the systematic generation of new examples of non-causal process functions.  
We also uncovered an unexpected bridge between causal and non-signaling inequalities. This 
construction turns out to be formally identical to that of Augusiak, Acín \textit{et al.}~%
\cite{augusiak11, augusiak12} for Bell inequalities without violation, their ``property~(P)'' 
coinciding with our unambiguity condition. Through this correspondence, every process function 
yields both a causal inequality and a non-signaling inequality of the local orthogonality 
type~\cite{fritz13}. Conversely, the nontrivial local orthogonality inequalities obtained from 
UPBs map to causal inequalities that no valid process function can maximally violate 
(Corollary~\ref{cor:upb}); their maximal violation requiring an invalid quasi-process, such as 
the global causal loop associated with the tripartite SHIFT UPB.

\medskip

We also note that while our analysis is restricted to process functions acting on finite-dimensional systems, process functions can also be defined for continuous variables~\cite{baumeler19}.   
Furthermore, while we focused here on deterministic cases, there also exist statistical and dynamical classical processes compatible with logical consistency that are non-causal~\cite{kunjwal23}.  
Such processes are expected to admit a correspondence with separable measurements\textemdash an extension we leave for future work.

\medskip

Our results provide a interpretative framework for   process functions and their possible non-causality formulated in terms of event labeling.
Unambiguity ensures that any pair of global events is pairwise exclusive, and that each local state $\ket{\psi^j_k}\in\mathcal{S}^{(k)}$ is uniquely and unambiguously labeled by a pair of classical output and input $(a_k^j|x_k^j)$.  
In contrast, the completeness of $\mathcal{S}$ guarantees that every global state $\ket{\psi^j}\in\mathcal{S}$, and its associated global event $(\bm{a}^j|\bm{x}^j)$, can be uniquely identified by the collection of classical outputs $\bm{a}^j$ alone.  
In other words, while local states require fine-grained labeling \textendash\ both $a_k^j$ and $x_k^j$ are in general necessary for their identification \textendash\ global states admit a coarse-grained labeling based solely on their output strings $\bm{a}^j$.  
Phrased differently, the identification of local events is setting-dependent, whereas the identification of the pairwise exclusive global events they compose is setting-independent.
Our results show that this event-labeling structure precisely captures logical consistency. This suggests a characterization of process functions solely based on their associated events list, that we formalize in a forthcoming work.  
Furthermore, non-causality and QNLWE arise exactly when every party (or when output-reduced subset of parties) can change their local basis. That is, they arise when no party's local events can be identified in a coarse-grained, setting-independent manner from outputs alone.  
In such cases, the fine-grained labeling remains essential for all parties, reflecting the non-causal structure of the underlying process.

\medskip

Finally, it was recently showed that the SHIFT basis can also be implemented using local operations with quantum control of classical communication (LOSupCC)~\cite{steffinlongo25}.  
Moreover, various Boolean process functions without a global past can be transformed into such a resource. 
This observation opens a pathway toward defining a hierarchy of deterministic classical communication classes, together with a corresponding hierarchy of unambiguous product bases, in direct analogy with the framework introduced in Ref.~\cite{wechs1}.  
Our results provide a natural foundation for such a program, by offering a principled generalization to arbitrary local dimensions.  
This, in turn, could shed new light on the operational advantages of indefinite causal order in information-processing tasks beyond the standard LOCC paradigm.

\begin{acknowledgments}
We thank Ognyan Oreshkov, Alastair Abbott, Cyril Branciard, Pierre Pocreau, Marco Túlio Quintino, Nasra Daher Ahmed, Alexei Grinbaum, Kuntal Sengupta and Ämin Baumeler for enlightening discussions. We are also grateful to Fabio Costa, who helped identify a loophole in an earlier version of this manuscript. We acknowledge the use of the AI assistant Claude (Anthropic), interaction with which helped identifying this loophole, as well as proving Appendix \ref{app:locbases}. The assistant was further used to improve the English of the manuscript. All mathematical statements and proofs were independently derived and are the sole responsibility of the authors. 

H.D., A.L., and A.A. acknowledge financial support from the Government of Spain (Severo Ochoa CEX2019-000910-S, NextGenerationEU PRTR-C17.I1 and FUNQIP), Fundació Cellex, Fundació Mir-Puig, Generalitat de Catalunya (CERCA program). A.A was also supported by the ERC AdG CERQUTE, the AXA Chair in Quantum Information Science. 
K.S., A.L., and E.C.B. acknowledge: This research was funded in whole or in part by the Austrian Science Fund (FWF) 10.55776/PAT4559623. For open access purposes, the author has applied a CC BY public copyright license to any author-accepted manuscript version arising from this submission. 
A.L. acknowledges the European Union (PASQuanS2.1, 101113690). 
R.K. acknowledges: This work received support from the French government under the France 2030 investment plan, as part of the Initiative d’Excellence d’Aix-Marseille Université-A*MIDEX, AMX-22-CEI-01.
R. A. and S. H. were supported by the European Union under Horizon Europe project NeQST (grant agreement no.~101080086). Views and opinions expressed are however those of the author(s) only and do not necessarily reflect those of the European Union or the European Commission. Neither the European Union nor the granting authority can be held responsible for them.
R. A. was also supported by the National Science Centre (Poland) through the SONATA BIS project No.
2019/34/E/ST2/00369. 
\end{acknowledgments}

\bibliography{bibli}

\newpage

\appendix

\section{Proof of Corollary \ref{cor:causalBound}}
\label{app:proofCausalBound}
Any non-causal process function $\w^{NC}$,
when combined with the  local strategy $P(a_k,o_k|x_k,i_k) = \delta_{a_k,i_k}\delta_{o_k,x_k}$, generates
non-causal correlations that maximally violate (with
probability one) the causal inequality~\eqref{eq:causalineq} associated with a causal game ``guess the process function $\w^{NC}$", with the success probability
\begin{equation}
    P_{\rm succ} := \frac{1}{\prod_k d_k}\sum_{\bm{i}} P(\bm{o}=\bm{w}(\bm{i}) | \bm{i}),
\end{equation}
where $d_k = |\mathcal{I}_k|$.

Assume that $P(\bm{o}|\bm{i})$ is an $n-$partite causal correlation in the sense of Eq.~\eqref{eq:causalcor}. The set of causal correlations is convex, and $P_{\rm succ}$ is linear in $P(\bm{o}|\bm{i})$. Hence the maximum of $P_{\rm succ}$ over all causal correlations is achieved by an extremal point, which corresponds to a deterministic strategy with a fixed causal order of the parties. 

Fix such a deterministic strategy and such a causal order. Then there exists a party $l$ that is not in the causal future of any other party. Hence, party $l$ cannot access $\bm{i}_{\backslash l}$ and its output $o_l$ can only depend on its local input $i_l$. On the other hand, the target value to be guessed is $x_l = w_l(\bm{i}_{\backslash l})$ which, for a nontrivial $\bm{w}$, is not constant as $\bm{i}_{\backslash l}$ varies. Therefore, there exists a value $i'\in\mathcal{I}_l$ such that, when restricting to inputs with $i_l = i'$, the values $w_l(\bm{i}_{\backslash l})$ take at least two distinct values as $\bm{i}_{\backslash l}$ ranges over its $\prod_{k\neq l} d_k$ possibilities. Since $o_l$ depends only on $i_l$, it is the same for all inputs with $i_l = i'$. Thus, among the $\prod_{k\neq l} d_k$ inputs with $i_l=i'$, party $l$ can be correct (i.e.\ $o_l=x_l$) on at most $\prod_{k\neq l} d_k - 1$ of them. 

In order to bound the success probability $P_{\rm succ}$, we split the sum according to whether $i_l=i'$ or not:
\begin{align}
    \nonumber P_{\rm succ} &= \frac{1}{\prod_k d_k} \Bigl( \sum_{\bm{i}: i_l\neq i'} P(\bm{o}= \bm{w}(\bm{i}) | \bm{i}) \\
    &\qquad + \sum_{\bm{i}: i_l= i'} P(\bm{o}= \bm{w}(\bm{i}) | \bm{i}) \Bigr).
\end{align}
For the first term, each probability is at most $1$, and there are $(d_l-1)\prod_{k\neq l} d_k$ such inputs because the local external input of party $l$ is fixed:
\begin{equation}
    \sum_{\bm{i}: i_l\neq i'} P(\bm{o}= \bm{w}(\bm{i}) | \bm{i}) \leq (d_l - 1)\prod_{k\neq l} d_k.
\end{equation}
For the second term, we restrict to inputs with $i_l = i'$. Among the $\prod_{k\neq l} d_k$ such inputs, at most $\prod_{k\neq l} d_k - 1$ can satisfy $\bm{o} = \bm{w}(\bm{i})$, as argued above. Therefore,
\begin{equation}
    \sum_{\bm{i}: i_l=i'} P(\bm{o}= \bm{w}(\bm{i}) | \bm{i}) \leq \prod_{k\neq l} d_k - 1.
\end{equation}
Combining both contributions, we obtain
\begin{align}
    P_{\rm succ}
    &\leq \frac{1}{\prod_k d_k}\left(
        (d_l-1)\prod_{k\neq l}d_k
        +
        \prod_{k\neq l} d_k - 1
    \right)\\
    &= \frac{1}{\prod_k d_k}\left(
        d_l \prod_{k\neq l} d_k - 1
    \right)\\
    &= \frac{1}{\prod_k d_k}\left(
        \prod_{k} d_k - 1
    \right)\\
    &= 1- \frac{1}{\prod_k d_k}.
\end{align}
Since this bound holds for every extremal causal strategy (and therefore for every causal correlation by convexity), the optimal value achievable by causal correlations is
\begin{equation}
    \beta^C = 1 - \frac{1}{\prod_k |\mathcal{I}_k|}.
\end{equation} 
In the special case $|\mathcal{I}_k| = d$ for all $k$, this reduces to $\beta^C = 1 - d^{-n}$, in agreement with Theorem~7 of Ref.~\cite{baumeler2}.

\section{Proof of Theorem \ref{thm:characcausalpf}}
\label{app:proofCausalPF}

\paragraph{\boldmath$(\Leftarrow)$}
We first show that any $n-$partite process function satisfying conditions ($i$) and ($ii$) of Theorem \ref{thm:characcausalpf} generates causal correlations in the sense of Eq.~\eqref{eq:causalcor}. The proof proceeds by induction on the number of parties.

\medskip

\emph{Base step $n=1$.} A single-party process function must be constant, $x_1 = w_1(a_1) := x_1^0$, by logical consistency. Substituting into Eq.~\eqref{eq:correlation2} gives
\begin{align}
    P(o_1|i_1) &= \sum_{a_1,x_1}  P(a_1,o_1|x_1,i_1)\delta_{x_1,x_1^0}, \\
    \label{eq:causalproof1-1} &= \sum_{a_1} P(a_1 | x_1^0,i_1)\, P(o_1 | x_1^0,i_1,a_1) \\
    \label{eq:causalproof1-2} &= \sum_{a_1}P(a_1 | x_1^0,i_1) P(o_1 | x_1^0,i_1),\\
    &= P(o_1 | x_1^0,i_1), \label{eq:causalproof1}
\end{align}
where~\eqref{eq:causalproof1-1} follows from the chain rule, and~\eqref{eq:causalproof1-2} from the fact that $o_1$ can only depend on $a_1$ via $x_1$, which is fixed. Defining $P_1(o_1|i_1):=P(o_1 | x_1^0,i_1)$, the resulting distribution is
trivially causal.

\medskip

\emph{Base step $n=2$.}
Assume without loss of generality that party $A_1$ is in the global past, so that $x_1 = x_1^0$ is constant. From Eq.~\eqref{eq:correlation2}, 
\begin{align}
    \nonumber P(\bm{o}|\bm{i})&= \sum_{\a,\x} P(a_1,o_1|x_1,i_1)P(a_2,o_2|x_2,i_2) \\
    &\qquad \cdot \delta_{x_1, x_1^0}\delta_{x_2,w_2(a_1)}\\
    \nonumber &= \sum_{a_1} P(a_1|x_1^0,i_1)P(o_1|x_1^0,i_1,a_1) \\
    &\qquad \cdot \sum_{a_2} P(a_2|w_2(a_1),i_2) P(o_2|w_2(a_1),i_2,a_2)\\
    \nonumber &= P(o_1|x_1^0,i_1) \\
    &\qquad \cdot \sum_{a_1} P(a_1|x_1^0,i_1) P(o_2|w_2(a_1),i_2) \\
    &:= P_1(o_1|i_1)P_{i_1}(o_2|i_2) \label{eq:causalproof2}
\end{align}
with $P_1(o_1|i_1) := P(o_1|x_1^0,i_1)$ and $P_{i_1}(o_2|i_2) := \sum_{a_1} P(a_1|x_1^0,i_1)P(o_2|w_2(a_1),i_2)$. This is precisely a bipartite causal correlation in the sense of Eq.~\eqref{eq:causalcor}. 

\medskip

\emph{Induction step.}
Assume that a $(n-1)-$partite process function satisfying $(i)$ and $(ii)$ generates causal correlations. We now show that this also holds for $n$. Let $\bm{w}$ be an $n-$partite process function satisfying ($i$) and ($ii$). By ($i$), there exists a party $l$ such that $x_l = x_l^0$ is constant. Using Eq.~\eqref{eq:correlation2} and separating the contribution of party $l$, we obtain 
\begin{align}
    P(\bm{o}|\bm{i})  
&=  \sum_{a_l, x_l} P(a_l, o_l|x_l,i_l) \delta_{x_l,x_l^0}  \notag\\ 
&\qquad \cdot \left[\sum_{\bm{a}_{\backslash l},\bm{x}_{\backslash l}} \prod_{k\neq l} P(a_k, o_k|x_k, i_k) \delta_{\bm{x}_{\backslash l}, \bm{w}^{a_l}(\bm{a}_{\backslash l})} \right]\\
&=  P(o_l|x_l^0,i_l) \sum_{a_l} P(a_l|x_l^0,i_l)  \notag\\ 
&\qquad \cdot \left[\sum_{\bm{a}_{\backslash l},\bm{x}_{\backslash l}} \prod_{k\neq l} P(a_k, o_k|x_k, i_k) \delta_{\bm{x}_{\backslash l}, \bm{w}^{a_l}(\bm{a}_{\backslash l})} \right]\\
    &:=  P_l(o_l|i_l)P_{i_l}(\bm{o}_{\backslash l}|\bm{i}_{\backslash l}),
\end{align}
where $P_l(o_l|i_l) := P(o_l|x_l^0,i_l)$, and
\begin{align}
\nonumber P_{i_l}(\bm{o}_{\backslash l}|\bm{i}_{\backslash l})
&:= \sum_{a_l} P(a_l|x_l^0,i_l) \\
&\qquad \cdot \sum_{\bm{a}_{\backslash l},\bm{x}_{\backslash l}} \prod_{k\neq l} P(a_k,o_k|x_k,i_k) \delta_{\bm{x}_{\backslash l},\bm{w}^{a_l}(\bm{a}_{\backslash l})}.
\end{align}
For each fixed value of $a_l$, the inner sum defines the correlation generated by the output-reduced process function $\bm{w}^{a_l} : \mathcal{A}_{\backslash l} \rightarrow \mathcal{X}_{\backslash l}$ in the sense of Eq.~\eqref{eq:output_red_pf}. By condition $(ii)$ for $\bm{w}$, each $\bm{w}^{a_l}$ is a valid $(n-1)-$partite causal process function. By the induction hypothesis, the corresponding correlations are causal for any fixed $a_l$. Since $P_{i_l}(\bm{o}_{\backslash l}|\bm{i}_{\backslash l})$ is a convex mixture of these causal correlations, it is itself causal. Hence $P(\bm{o}|\bm{i})$ admits the decomposition required by Eq.~\eqref{eq:causalcor}, and is therefore an $n-$partite causal correlation.

\medskip

\paragraph{\boldmath$(\Rightarrow)$}
We now prove the converse direction, namely that any $n-$partite process function $\bm{w}:\mathcal{A}\to\mathcal{X}$ that generates only causal correlations satisfies conditions $(i)$ and $(ii)$ of Theorem~\ref{thm:characcausalpf}. The proof proceeds by induction on $n$.

\medskip

\emph{Base step $n=1$.}
For a single party, logical consistency implies that the process function must be constant, $x_1 = w_1(a_1) := x_1^0$. Hence condition $(i)$ is satisfied, and $(ii)$ is trivial.

\medskip

\emph{Base step $n=2$.}
Any bipartite causal correlation is compatible with a definite causal order. Since we consider deterministic process functions, this means that either $A_1 \prec A_2$ or $A_2 \prec A_1$, where $A_i$ denotes party $i$. Without loss of generality, assume $A_1 \prec A_2$, so that
\begin{equation}
P(o_1,o_2|i_1,i_2) = P(o_1|i_1)\,P_{1,i_1,o_1}(o_2|i_2).
\end{equation}
From the characterization of bipartite process functions, this implies that $w_1$ is constant, establishing $(i)$. Moreover, fixing any output $a_1$ yields a single-party output-reduced process function $\bm{w}^{a_1}$, which is necessarily constant and thus causal, establishing $(ii)$. 

\medskip

\emph{Induction step.}
Assume that the statement holds for a $(n-1)-$partite process function. Let $\bm{w}$ be an $n-$partite process function that generates only causal correlations. Then, by definition of multipartite causal correlations~\eqref{eq:causalcor}, the resulting distribution can be written as
\begin{equation}
    P(\bm{o}|\bm{i}) = P(o_1|i_1)\,P_{1,i_1,o_1}(\bm{o}_{\backslash 1}|\bm{i}_{\backslash 1}),
\end{equation}
where party $A_1$ is in the global causal past of the remaining parties. Since no other party can signal to $A_1$, the input $x_1$ of the process function cannot depend on $\bm{a}_{\backslash 1}$, and hence $w_1:\mathcal{A}_{\backslash 1} \rightarrow \mathcal{X}_1$ is constant. This establishes condition $(i)$. 

Furthermore, by definition of causal correlations,
$P_{1,i_1,o_1}(\bm{o}_{\backslash 1}|\bm{i}_{\backslash 1})$ is a causal $(n-1)-$partite correlation for every pair $(i_1,o_1)$. These correlations are in general by the output-reduced process functions $\bm{w}^{a_1}$, obtained by fixing the local operation of party $A_1$. By the induction hypothesis, all output-reduced process functions $\bm{w}^{a_1}$ are causal, and thus satisfy $(ii)$. This establishes condition $(ii)$ for $\bm{w}$. This completes the induction and the proof.

\section{Proof of Lemma \ref{lem:PE}}\label{app:lemmaPE}
    Let $\mathcal{S}$ be unambiguous. Then $\mathcal{S}$ satisfies pairwise exclusivity: for any distinct $|\psi^j\rangle, |\psi^{j'}\rangle \in \mathcal{S}$, there exists $k \in \{1, \ldots, n\}$ such that $x_k^j = x_k^{j'}$ and $a_k^j \neq a_k^{j'}$.
    \begin{proof}
        Let $|\psi^j\rangle, |\psi^{j'}\rangle \in \mathcal{S}$ with $j\neq j'$. Since $\mathcal{S}$ is an orthonormal basis, we obtain:
        \begin{equation}
           \langle\psi^j|\psi^{j'}\rangle = \prod_{k=1}^n \langle{(a_k^j| x_k^j)}|(a_k^{j'}|x_k^{j'})\rangle=0.
        \end{equation}
        Therefore, there exists $k \in \{1, \ldots, n\}$ such that 
        \begin{equation}\label{app:eq:orthLoc}
            \langle(a_k^j|x_k^j)|(a_k^{j'}|x_k^{j'})\rangle = 0.    
        \end{equation}
        Due to~\eqref{def:UA}, local vector are orthogonal only if their settings match:
        \begin{equation}
            x_k^j = x_k^{j'}.
        \end{equation}
        On the other hand,~\eqref{def:UA} guarantees that, for any $|(a_k|x_k)\rangle, |(a'_k|x_k)\rangle \in \mathcal{S}_{x_k}^{(k)}$ for some $x_k$ and $a_k \neq a_k'$, we have
        \begin{equation}
            \langle(a_k^j| x_k^j)|(a_k^{j'}|x_k^{j})\rangle = 0,
        \end{equation}
        which, together with completeness of $\mathcal{S}$ means that $\mathcal{S}_{x_k}^{(k)}$ is an orthonormal basis. Therefore,~\eqref{app:eq:orthLoc} implies that $a_k^j \neq a_k^{j'}$, proving the thesis.
    \end{proof}

\section{Completeness and unambiguity implies complete local bases}\label{app:locbases}

\begin{theorem}
Let $\mathcal H=\bigotimes_{k=1}^n\mathbb C^{d_k}$ and let 
$\mathcal S=\{\ket{\psi^j}=\ket{\psi^j_1}\otimes\cdots\otimes \ket{\psi^j_n}\}_{j=1}^{|\mathcal S|}$ 
be an unambiguous complete product basis, i.e.\ a set of mutually orthogonal product 
vectors with $|\mathcal S|=\prod_{k=1}^n d_k$ such that, for every party $k$, two distinct 
local vectors in the local set $\mathcal{S}^{(k)} = \bigsqcup_{x_k=0}^{|\mathcal{X}_k|-1} \mathcal{S}_{x_k}^{(k)}$ are orthogonal if and only if they belong to the same local subset 
$\mathcal S^{(k)}_{x_k}$. Then, for every party $k$ and every setting $x_k\in\mathcal X_k$, 
the subset $\mathcal S^{(k)}_{x_k}$ is a complete orthonormal basis of $\mathbb C^{d_k}$; 
in particular $|\mathcal A_k|=d_k$ and every setting $x_k$ is a complete rank-one measurement.
\end{theorem}

\begin{proof}
Fix a party $k$ and a setting $x_k$. By unambiguity, the vectors of 
$\mathcal S^{(k)}_{x_k}$ are mutually orthogonal, so
\begin{align}
    |\mathcal S^{(k)}_{x_k}|\le d_k
\end{align} and 
they span a subspace $V:=\operatorname{span}\mathcal S^{(k)}_{x_k}\subseteq\mathbb C^{d_k}$ 
of dimension $|\mathcal S^{(k)}_{x_k}|$. It remains to show that $V=\mathbb C^{d_k}$.

Assume, for contradiction, that $V\subsetneq\mathbb C^{d_k}$, and pick a unit vector 
$\ket\phi\in V^\perp$. Let $\ket{\psi^{j_0}}=\ket{\psi^{j_0}_k}\otimes\ket{\psi^{j_0}_{\bar k}}\in\mathcal S$ 
be any basis element whose $k$-th component $\ket{\psi^{j_0}_k}$ lies in 
$\mathcal S^{(k)}_{x_k}$, where 
$\ket{\psi^{j_0}_{\bar k}}\in\mathcal H_{\bar k}:=\bigotimes_{l\neq k}\mathbb C^{d_l}$ 
denotes its component on the remaining parties.

Consider the unit vector $\ket\Phi:=\ket\phi\otimes\ket{\psi^{j_0}_{\bar k}}$. Expanding it 
in the orthonormal basis $\mathcal S$ (Parseval's identity),
\begin{equation}
    1=\|\ket\Phi\|^2=\sum_j\big|\braket{\psi^j}{\Phi}\big|^2
    =\sum_j\big|\braket{\psi^j_k}{\phi}\big|^2\,
     \big|\braket{\psi^j_{\bar k}}{\psi^{j_0}_{\bar k}}\big|^2 .
\end{equation}
The sum being nonzero, there exists an index $j^*$ for which both factors are nonzero, i.e.
\begin{equation}
    \braket{\psi^{j^*}_k}{\phi}\neq0
    \qquad\text{and}\qquad
    \braket{\psi^{j^*}_{\bar k}}{\psi^{j_0}_{\bar k}}\neq0 .
\end{equation}
Since $\ket\phi\in V^\perp$, the condition $\braket{\psi^{j^*}_k}{\phi}\neq0$ forces 
$\ket{\psi^{j^*}_k}\notin V$; in particular $\ket{\psi^{j^*}_k}\notin\mathcal S^{(k)}_{x_k}$ 
and $\ket{\psi^{j^*}_k}\neq\ket{\psi^{j_0}_k}$. Thus $\ket{\psi^{j_0}_k}$ and 
$\ket{\psi^{j^*}_k}$ are distinct local vectors belonging to different subsets, so 
unambiguity gives $\braket{\psi^{j_0}_k}{\psi^{j^*}_k}\neq0$. Combining this with 
$\braket{\psi^{j^*}_{\bar k}}{\psi^{j_0}_{\bar k}}\neq0$,
\begin{equation}
    \braket{\psi^{j_0}}{\psi^{j^*}}
    =\braket{\psi^{j_0}_k}{\psi^{j^*}_k}\,
     \braket{\psi^{j_0}_{\bar k}}{\psi^{j^*}_{\bar k}}\neq0 ,
\end{equation}
with $j_0\neq j^*$, contradicting the mutual orthogonality of $\mathcal S$.

Hence $|\mathcal S^{(k)}_{x_k}|=d_k$. A set of $d_k$ mutually orthogonal unit vectors in 
$\mathbb C^{d_k}$ is a complete orthonormal basis, so each $\mathcal S^{(k)}_{x_k}$ is an 
orthonormal basis of $\mathbb C^{d_k}$, and every setting $x_k$ is a complete rank-one 
measurement with $|\mathcal A_k|=d_k$.
\end{proof}

\section{Auxiliary Lemmas for Theorem \ref{thm:unamb_to_pf}}
\label{app:lemmasUA}

Let $\mathcal{S} = \{ |\psi^j\rangle \}_{j=1}^{|\mathcal S|} \subseteq \bigotimes_{k=1}^n \mathbb{C}^{d_k}$ be a complete $n-$partite orthonormal product basis, where $|\psi^j\rangle = \bigotimes_{k=1}^n |\psi_k^j\rangle$ and $|\mathcal S|=\prod_{k=1}^n d_k$. In this Appendix (as well as in Appendix \ref{app:weakUA}), $\S^{k}$ and $\S_{x_k}^{k}$ are not directly defined as subsets of $\S$, but as tuples of states associated to local parties, mapped into the global states of $\S$ (see also Footnote \ref{fnt:tuple}).

\medskip

For each party $k$, let $\mathcal X_k$ be a finite set of local settings and $\mathcal A_k=\{0, \ldots, d_k-1\}$ outcomes, and let the local labeled family be given by fixed labels 
\begin{equation}
    (a_k|x_k)\longmapsto |(a_k|x_k)\rangle\in\mathbb C^{d_k}, \quad x_k\in\mathcal X_k,\ a_k\in\mathcal A_k.
\end{equation}

Define the tuples of local states \footnote{\label{fnt:tuple} In this Appendix, we adopt a generalized definition of local set $\mathcal S^{(k)}$ as a (not necessarily disjoint) union $\bigcup_{x_k\in\mathcal X_k} S^{(k)}_{x_k}$ of settings, which can share local vectors. For any unambiguous $\mathcal S$, it reduces to a disjoint union $\bigsqcup_{x_k\in\mathcal X_k} S^{(k)}_{x_k}$ of Section \ref{sec:upbpf}, as shown in Lemma \ref{app:lemma:labellingUA}.}
\begin{equation}
\mathcal S^{(k)}=\bigcup_{x_k\in\mathcal X_k}\ \mathcal S^{(k)}_{x_k},
\qquad
\mathcal S^{(k)}_{x_k}=\{|(a_k|x_k)\rangle : a_k\in\mathcal A_k\}
\end{equation}
such that there exists a surjective map
\begin{equation}
    \bigtimes_{k=1}^n |(a_k|x_k)\rangle \mapsto \bigotimes_{k=1}^n |(a_k|x_k)\rangle \in \mathcal{S},
\end{equation}
for any $|(a_k|x_k)\rangle \in \mathcal{S}^{(k)}$.

First, we prove several useful properties of unambiguity that is defined in~\eqref{def:UA}.

\begin{lemma}\label{app:lemma:labellingUA}
    Let $\mathcal{S}$ be unambiguous. Then, for any $k \in \{1, \ldots, n\}$ and $x_k \neq x'_k$,
    \begin{equation}
        \mathcal{S}^{(k)}_{x_k} \cap \mathcal{S}^{(k)}_{x'_k} = \emptyset.
    \end{equation}
    \begin{proof}
        Assume, for contradiction, that there exist $x_k \neq x'_k$ such that $\mathcal{S}^{(k)}_{x_k} \cap \mathcal{S}^{(k)}_{x'_k}$ is non-empty. Let $|\phi\rangle \in \mathcal{S}^{(k)}_{x_k} \cap \mathcal{S}^{(k)}_{x'_k}$, so that
        \begin{equation}
            |\phi\rangle = |(a_k|x_k)\rangle = |(a'_k|x'_k)\rangle,
        \end{equation}
        for some $a_k, a'_k \in \mathcal{A}_k$. Within a fixed setting $x_k$,~\eqref{def:UA} ensures that distinct vectors are orthogonal. Hence, for any $|(b_k|x_k)\rangle \in \mathcal{S}_{x_k}^{(k)}$ with $b_k \neq a_k$,
        \begin{equation}
            \langle (b_k|x_k) | \phi \rangle = 0,
        \end{equation}
        because $|\phi\rangle \in \mathcal{S}_{x_k}^{(k)}$. But since the same vector $| \phi \rangle$ lies in $\mathcal{S}_{x'_k}^{(k)}$, we conclude that $\mathcal{S}_{x_k}^{(k)} \subseteq \mathcal{S}_{x'_k}^{(k)}$. By symmetry, we obtain the inverse inclusion $\mathcal{S}_{x'_k}^{(k)} \subseteq \mathcal{S}_{x_k}^{(k)}$ as well, so that
        \begin{equation}
            \mathcal{S}_{x_k}^{(k)} = \mathcal{S}_{x'_k}^{(k)}.
        \end{equation}
        Therefore, the assumption that $x_k$ and $x'_k$ are distinct setting labels is wrong, and we conclude that the intersection of $\mathcal{S}_{x'_k}^{(k)}$ and $\mathcal{S}_{x_k}^{(k)}$ must be empty.
    \end{proof}
\end{lemma}

We proceed by showing that unambiguity of $\mathcal{S}$ provides existence of a unique map between joint outcomes and joint settings in the following Lemma.

\begin{lemma}\label{app:lemma:funcBasis}
    Let $\mathcal{S}$ be unambiguous. Then there exists a unique function $\bm{w}:\mathcal A\to\mathcal X$ such that
    \begin{equation}\label{app:eq:setW}
        \mathbf{W}(\mathcal S) = \{(\bm{a}|\bm{x}) : \bm{x} = \bm{w}(\bm{a})\} \subseteq \mathcal A\times\mathcal X,     
    \end{equation}
    where $\mathcal X=\bigtimes_k \mathcal X_k$ and $\mathcal A=\bigtimes_k \mathcal A_k$. 
\begin{proof}
    For each party $k$, consider the labelled local set $\mathcal{S}^{(k)}$ where the labels $(x_k,a_k)$ are assumed fixed in advance. Define the local evaluation map that associates an event to local state 
    \begin{equation}
        \mathbf{E}_k : \mathcal{S}^{(k)} \to \mathcal A_k\times \mathcal X_k, \qquad \mathbf{E}_k(|(a_k|x_k)\rangle)=(a_k | x_k),
    \end{equation}
    which is injective by construction of the labels. Set $\mathbf{E} := \prod_{k=1}^n \mathbf{E}_k$.

    \medskip

    Next, define the local-factor map
    \begin{equation}
        \mathbf{L}:\ \mathcal S \longrightarrow \bigtimes_{k=1}^n \mathcal S^{(k)}, \qquad \mathbf{L}(|\psi^j\rangle) = \{|(a_k^j|x_k^j)\rangle\}_{k=1}^n,
    \end{equation}
    where $(a_k^j|x_k^j)$ is the unique label of $|\psi_j^{(k)}\rangle$. By Lemma \ref{app:lemma:labellingUA}, every local vector of every basis element appears with a unique setting–outcome label, and therefore no two distinct global basis vectors share the same local vector. Hence $\mathbf{L}$ is injective.

       \medskip

    Define now the global event map
    \begin{equation}
        \mathbf{W} := \mathbf{E}\circ\mathbf{L}:\ \mathcal S \to \mathcal A\times\mathcal X, \qquad \mathbf{W}(|\psi^j\rangle)=(\bm{a}^j|\bm{x}^j),
    \end{equation}
    which is injective, since both $\mathbf{E}$ and $\mathbf{L}$ are injective. Consider an arbitrary joint outcome $\bm a \in \mathcal A$ and the corresponding fiber
    \begin{equation}
        \mathcal F(\bm{a}):=\{\bm{x}\!: (\bm{a}| \bm{x})\in \mathbf{W}(\mathcal S)\}.
    \end{equation}
    Suppose, \textit{ad absurdum}, that there exist two distinct vectors $|\psi^{j'}\rangle\neq |\psi^{j}\rangle$  in $\mathcal S$, both associated with the same joint outcome $\a$ but with different joint settings $\bm{x'} \neq \bm{x}$. Then, by Lemma \ref{lem:PE}, there exists a party $k$ at which the local vectors $|\psi^j_k\rangle$ and $|\psi_k^{j'}\rangle$ share the same setting $x_k = x_k'$ but different outcomes $a_k\neq a_k'$, which contradicts that the two events have identical joint outcome $\bm{a}$. Therefore, each $\mathcal F(\bm{a})$ contains at most one element, that is, $|\mathcal F(\bm{a})|\le 1$.

    \medskip

    Because $\mathcal S$ is a complete orthonormal product basis, we have $|\mathcal S|=\prod_k d_k=|\mathcal A|$, and since $\mathbf{W}$ is injective,
    \begin{equation}
        |\mathbf{W}(\mathcal S)| = \sum_{\bm{a}\in\mathcal A} |\mathcal F(\bm{a})| = |\mathcal A|.
    \end{equation}
    With every fiber having size at most one and the sum equal to $|\mathcal A|$, it follows that
    \begin{equation}
        |\mathcal F(\bm{a})|=1 \qquad \forall \bm{a}\in\mathcal A.
    \end{equation}
    Thus, for every joint outcome $\bm{a}$, there exists a unique joint setting $\bm{x}$ such that $(\bm{a}|\bm{x})\in\mathbf{W}(\mathcal S)$. We therefore define the map
    \begin{equation}
        \bm{w} : \mathcal A \to \mathcal X,\qquad \bm{w}(\bm{a}) = \bm{x}.
    \end{equation}
    Uniqueness of $\bm{w}$ follows immediately from the uniqueness of each fiber.

\end{proof}
\end{lemma}

\section{Unambiguity and weak unambiguity}
\label{app:weakUA}

Although we adopt~\eqref{def:UA} as the working definition of unambiguity, it can be relaxed by requiring only the structural properties established in Lemmas \ref{app:lemma:labellingUA} and \ref{lem:PE}. This leads to a weaker notion introduced below.
\begin{definition}[Weak unambiguity]
    Let $\mathcal{S}$ be a complete $n-$partite orthonormal product basis. It is called \textit{weakly unambiguous} if: (i) for any  $k \in \{1, \ldots, n\}$ and any two distinct settings $x_k \neq x'_k$,
    \begin{equation}
        \mathcal{S}_{x_k}^{(k)} \cap \mathcal{S}_{x'_k}^{(k)} = \emptyset,
    \end{equation}
    and, (ii) for any distinct basis vectors $|\psi^j\rangle, |\psi^{j'}\rangle \in \mathcal{S}$, there exists $k \in \{1, \ldots, n\}$ such that 
    \begin{equation}
        x_k^j = x_k^{j'}, \qquad a_k^j \neq a_k^{j'}.
    \end{equation}
\end{definition}
By construction, every unambiguous basis in the sense of~\eqref{def:UA} is weakly unambiguous, and every weakly unambiguous basis $\mathcal{S}$ determines a process function $\bm{w}$. However, the converse does not hold: not every weakly unambiguous basis is unambiguous. Consider the bipartite basis:
\begin{eqnarray}
\ket{(0|0)} \ket{(0|0)} &=&  \ket{0}\ket{0}, \\
\ket{(0|0)} \ket{(1|0)} &=& \ket{0}\ket{1}, \\
\ket{(0|0)} \ket{(2|0)} &=&  \ket{0}\ket{2}, \\
\ket{(1|0)} \ket{(0|1)} &=&  \ket{1}\ket{0+1+2}, \\
\ket{(1|0)} \ket{(1|1)} &=&  \ket{1}\ket{0-1}, \\
\ket{(1|0)} \ket{(2|1)} &=&  \ket{1}(\ket{0+1}-2\ket{2}),
\end{eqnarray}
where $\ket{0+1+2} = \frac{1}{\sqrt{3}}(\ket{0} + \ket{1} + \ket{2})$, and $\ket{0\pm1} = \frac{1}{\sqrt{2}}(\ket{0} \pm \ket{1})$. The corresponding local settings are
\begin{eqnarray}
    \mathcal{S}_0^{(1)} &=& \{\ket{0}, \ket{1}, \ket{2}\}, \\
    \mathcal{S}_0^{(2)} &=& \{\ket{0}, \ket{1}, \ket{2}\}, \\
    \mathcal{S}_1^{(2)} &=& \{\ket{0+1+2}, \ket{0-1}, \ket{0+1}-2\ket{2}\}.
\end{eqnarray}
The local settings within each party are non-overlapping and satisfy the disjointness condition of Lemma \ref{app:lemma:labellingUA}, and pairwise exclusivity (Lemma \ref{lem:PE}) can be verified directly by inspection of the indices $x_k$ and $a_k$. Nevertheless, it is not unambiguous in the sense of~\eqref{def:UA}, because some vectors belonging to different local settings are accidentally orthogonal, e.g., $|2\rangle \in \mathcal{S}_0^{(2)}$ and $|0-1\rangle \in \mathcal{S}_1^{(2)}$.

Despite this distinction, it suffices to work with unambiguous bases. The following Lemma shows that any weakly unambiguous basis can be made unambiguous by suitable local rotations, without altering the corresponding process function.

\begin{lemma}
    Let $\mathcal{S}$ be weakly unambiguous but not unambiguous, and let $\bm{w}$ denote its associated process function.  Then there exists an unambiguous basis $\mathcal{S}'$ with process function $\bm{w}'$ such that
    \begin{equation}
        \bm{w}' = \bm{w}.
    \end{equation}
    \begin{proof}

        If $\mathcal{S}$ is weakly unambiguous but not unambiguous, then there exists $k \in \{1, \ldots, n\}$ and distinct settings $x_k \neq x_k'$ for which
        \begin{equation}
            \langle (a_k| x_k) | (a'_k | x'_k) \rangle = 0
        \end{equation}
        for some $a_k, a_k' \in \mathcal{A}_k$. Since each $\mathcal{S}_{x_k}^{(k)}$ is an orthonormal basis on $\mathbb{C}^{d_k}$, one may redefine it via a local unitary $V_{x_k}^{(k)} \in \mathbf{U}(d_k)$ such that
        \begin{equation}
            |(a_k|x_k)\rangle \mapsto  V_{x_k}^{(k)}|(a_k|x_k)\rangle.
        \end{equation}
        chosen so that overlaps between vectors belonging to different settings become nonzero. Such unitaries exist generically since the set of $V_{x_k}^{(k)}$ leaving any given overlap between local vectors from different settings zero has measure zero in $\mathbf{U}(d_k)$. These local transformations preserve orthonormality within each setting, the product structure of the global basis $\mathcal{S}$, and all event labels $(a_k|x_k)$. Consequently, the set $\mathbf{W}(\mathcal{S})$ and its induced process function remain unchanged, yielding an unambiguous basis $\mathcal{S}'$ with process function $\bm{w}' = \bm{w}$.
    \end{proof}
\end{lemma}

\section{Different ambiguous product bases and their connection to process functions}
\label{app:ambi}

In this section, we examine various cases of ambiguous product bases and their relation to process functions.

\subsection{Ambiguous QNLWE product basis generating a non-causal quasi-process function}\label{app:domino}

The canonical example of QNLWE is known as the domino basis \cite{bennett_99}, the two-qutrit QNLWE basis\begin{align}
    \{\ket{0+},\ket{0-},\ket{+2},\ket{+'0},\ket{11},\ket{-2},\ket{-'0},\ket{2+'},\ket{2-'}\}
\end{align}
The associated local operations are projections in $\{\ket{0},\ket{1},\ket{2}\}_{x_k=0} $, $\{\ket{+},\ket{-},\ket{2}\}_{x_k=1}$ or $\{\ket{0},\ket{+'},\ket{-'}\}_{x_k=2}$ for both Alice and Bob, with $\ket{\pm}=(\ket{0}\pm\ket{1})\sqrt{2}$ and $\ket{\pm'}=(\ket{1}\pm\ket{2})\sqrt{2}$. This example does not satisfy unambiguity~\eqref{def:UA}. Therefore, the local label $(a_k|x_k)$ of both $\ket{0}$ and $\ket{2}$ is ambiguous. \\

This labeling ambiguity can lead to a violation of the pairwise exclusivity of events~\eqref{eq:PE}, despite each state being orthogonal.  Consider for instance the two states $\ket{11}$ and $\ket{+2}$. They are orthogonal, since $\braket{1}{2}=0$ in the computational basis. The associated events, written in the form $(a_1a_2|x_1x_2)$, should be $(11|00)$ and $(02|10)$ respectively, which are orthogonal\textemdash i.e.~they share at least one common local input, with distinct outputs assigned to it. However, the event associated with $\ket{+2}$ could  actually also be $(02|11)$, which is not orthogonal to  $(11|00)$. This ambiguity arises because the state $\ket{2}$ can be obtained in two different measurement contexts.

In fact, one can verify that there is no globally consistent assignment of events for all the states in the domino basis that would satisfy Eq.\eqref{eq:PE}: for example, $\ket{11}$ enforces the label $(02|10)$ for $\ket{+2}$, while $\ket{0+}$ enforces the label $(02|11)$, leading to a contradiction.

Nevertheless, it is possible to construct a non-valid quasi-process function from this basis: the two-way-signaling process  $(x_1:=a_2\oplus_3 2,x_2:=a_1\oplus_{(2)} 1)$, where $\oplus_3$ is the addition modulo 3, and $\oplus_{(2)}$ the operation that flips 0 and 1 into one another while leaving 2 unchanged. To verify this, one can retrieve the domino basis from this non-valid process using the local unitaries $U^{x_k=0}=\id$, $U^{x_k=1}=H\oplus\id $ (Hadamard on the $\{\ket{0},\ket{1}\}$ subspace) and $U^{x_k=2}=\id\oplus H $ (Hadamard on the $\{\ket{1},\ket{2}\}$ subspace) for both parties, 
which, when applied to the computational basis, correspond to projections in the corresponding local bases. Note that under unitaries satisfying the unambiguity constraint Eq.\eqref{eq:unambunit}, the process  no longer generates a valid basis.

\subsection{Ambiguous product basis generating a valid process function}
\label{app:amb_valid}
Consider the following product basis in $2\otimes3$.
\begin{align}\label{product1}
\{\ket{00},~\ket{01},~\ket{02}, ~\ket{1+},~\ket{1-},~\ket{12}\},
\end{align}
where the qubit belongs to Alice and the qutrit belongs to Bob. Alice performs a local projection in $\{\ket{0},\ket{1}\}_{x_1=0}$, while Bob has two choices: $\{\ket{0},\ket{1},\ket{2}\}_{x_2=0}$ and $\{\ket{+},\ket{-},\ket{2}\}_{x_2=1}$. The product basis is ambiguous, as $\ket{2}$ appears in both bases $x_2=0$ and $x_2=1$. Thus unambiguity (Eq.\eqref{def:UA}) is not satisfied. The associated events $(a_1a_2|x_1x_2)$ are:

\begin{equation}\label{labeling_product1}
\begin{array}{ll}
\ket{00}\rightarrow(00|00), & \ket{1+}\rightarrow(10|01),\\[1 ex]

\ket{01}\rightarrow(01|00), & \ket{1-}\rightarrow(11|01),\\[1 ex]

\ket{02}\rightarrow(02|00)/(02|01), & \ket{12}\rightarrow(12|01)/(12|00).\\
\end{array}
\end{equation}
Unlike the domino example, here pairwise exclusivity of events~\eqref{eq:PE} is always satisfied, and a specific labeling $(\ket{02}\rightarrow(02|00))$, $(\ket{12}\rightarrow(12|01))$ give rise to a valid process function, $(x_1=0,x_2=a_1)$, which correspond to a one-way signaling from Alice to Bob. However, the other possible labels do not give a valid function $x_2=w_2(a_1)$. 

\subsection{Ambiguous local product basis generating a non-causal quasi-process function}\label{app:amb_non-causal}

Consider another product basis in $2\otimes3$, given as the following:
\begin{equation}\label{product3}
\{\ket{00},~\ket{01},~\ket{1+},~\ket{1-},
\ket{+2},~\ket{-2}\}
\end{equation}
where on both sides there are more than one local bases. On Alice's side, $\{\ket{0}, \ket{1}\}_{x_1=0}$ and $\{\ket{+}, \ket{-}\}_{x_1=1}$. On Bob's side, $\{\ket{0}, \ket{1}, \ket{2}\}_{x_2=0}$ and $\{\ket{+}, \ket{-}, \ket{2}\}_{x_2=1}$. The basis is clearly ambiguous due to the presence of $\ket{2}$ in Bob's two local bases. The ambiguity give rise to a violation of Eq.   ~\eqref{eq:PE}, e.g.\ with the labeling $\ket{1+}\rightarrow (10|01)$ and $\ket{+2}\rightarrow (02|10)$: despite the states being orthogonal, the associated events are not.  Therefore, this basis does not correspond to a valid process function. It can nevertheless be associated with the non-causal quasi-process function $(x_1=\delta_{a_2,2},x_2=a_1)$, corresponding to a two-way-signaling between Alice and Bob. Let us recall that two-way signaling is a form of non-causality forbidden in single round scenarios, as it can give rise to logical antinomies. Thus this basis cannot be discriminated by single-round LOCC, as it involves two-way signaling. However, unlike the domino basis, the basis Eq.\eqref{product3} is not QNLWE, as it can be measured by multi-round LOCC, following a strategy presented in \cite{bennett_99}: Bob initiates the protocol with an incomplete von-Neumann measurement distinguishing his state $\ket{2}$ from $\ket{0}$ and $\ket{1}$ (without distinguishing $\ket{0}$ from $\ket{1}$). If $a_2=2$, Alice then measures in the diagonal basis, discriminating states $\ket{+2}$ and $\ket{-2}$ ; if $a_2\neq 2$, she measures in the computational basis, and sends the outcome to Bob. If $a_1=0$, Bob measures in the computational basis and discriminates states $\ket{00}$ and $\ket{01}$ ; $a_1=1$, Bob measures in the diagonal basis and discriminates states $\ket{1+}$ and $\ket{1-}$.  

\subsection{Discussion}\label{app:amb_conclu}

As unambiguity is always satisfied with qubits, all qubit product bases are associated with valid process functions. In the qudit case, unambiguity is not systematic. When it is not satisfied, various examples arise: some ambiguous product bases which exhibit QNLWE are never associated with valid process functions, such as the domino basis (cf.~\ref{app:domino}). Some ambiguous product bases that do not exhibit QNLWE can admit a label with locally orthogonal events associated with a valid process function (cf.~\ref{app:amb_valid}), while others cannot (cf.~\ref{app:amb_non-causal}). In the latter case, the basis is not locally measurable by single-round LOCC nor LOPF, but can nevertheless be measured by multi-round LOCC: in other words, bases of this form could be deemed ``single-round non-local''. 
We leave a more general and systematic study of these instances for future work.

\section{LOPF defines a D-PVM}\label{app:proofth1}

To show that local projective measurements implemented via a process function give rise to an effective distributed projective measurement (D-PVM)~\cite{supic17,hoban18,dourdent21,dourdent24}, we first establish a more general result. Specifically, we show that local quantum instruments~\cite{davies70} with quantum inputs, when implemented within process matrices~\cite{oreshkov1}\textemdash of which process functions constitute a special case~\cite{baumeler22}\textemdash generate an effective distributed measurement (D-POVM).

\medskip 

Process functions are the deterministic classical counterpart of
process matrices~\cite{oreshkov1}, which are the Choi representations~\cite{choi75} of quantum supermaps that yield valid probabilities under arbitrary quantum operations.\medskip

In the standard formalism, each party $k$ performs a single local quantum operation 
\begin{equation}
    M_{|i_k}: \mathcal{L}(\HS^{A^{I'}_k}) \times \mathcal{I}_k \rightarrow \mathcal{L}(\HS^{A^{O'}_k}) \times \mathcal{O}_k,\label{eq:instru}
\end{equation}

where, compared to Eq.~\eqref{eq:localpovm}, the classical spaces $\X_k$ and $\A_k$ are replaced by spaces of linear operators $\mathcal{L}(\HS^{A^{I'}_k})$ and $\mathcal{L}(\HS^{A^{O'}_k})$. Concretely, $M_{|i_k}$ is the Choi operator~\cite{choi75} of a quantum instrument
$M_{|i_k} := (M_{o_k|i_k}^{A^{I'}_kA^{O'}_k})_{o_k|i_k}$, where each element 
$M_{o_k|i_k}^{A^{I'}_kA^{O'}_k} \in \mathcal{L}(\HS^{A^{I'}_kA^{O'}_k}):=\mathcal{L}(\HS^{A^{I'}_k}\otimes\HS^{A^{O'}_k})$ for any $i_k\in\I_k$ satisfies
$M_{o_k|i_k}^{A^{I'}_kA^{O'}_k} \ge 0$ and 
$\Tr_{A^{O'}_k} \sum_{o_k} M_{o_k|i_k}^{A^{I'}_kA^{O'}_k} = \openone^{A^{I'}_k}$. \\

With the notations  $\mathbf{A}^{'} = A_1^{'}\ldots A_n^{'}$, $A_k' = A_k^{I'} A_k^{O'}$,  $\bm{o} = (o_1, \ldots, o_n)$ and $\bm{i} = (i_1, \ldots, i_n)$,
within the process matrix framework, the correlations established by the parties are given by the probabilities
\begin{align}
    P(\bm{o}|\bm{i})&=\Tr\left[
\big(\bigotimes_{k=1}^n M_{o_k|i_k}^{A'_k}\big)^T\cdot
W^\mathbf{A'}
\right],\notag\\
&=\bigotimes_{k=1}^n M_{o_k|i_k}^{A'_k}*W^\mathbf{A'},\label{eq:probapm0}
\end{align}
where $W\in\mathcal{L}(\HS^{\mathbf{A'}})$ is the ``process matrix", and $*$ denotes the ``link product''~\cite{chiribella08,chiribella09}. For two operators $M\in\L(\mathcal{H}^{XY})$ and $N\in\L(\mathcal{H}^{YZ})$, the latter is defined as $M*N=\Tr_{Y} [(M^{T_Y}\otimes\id^{Z})(\id^{X}\otimes N)] $ where $T_Y$ denotes  partial transposition on  $Y$. This product is both commutative (up to reordering of tensor products) and associative (as long as each space involved in the product appears at most twice) ; a tensor product $M^X\otimes N^Y$ is a special case with trivial link.

\medskip

For Eq.~\eqref{eq:probapm0} to define valid probabilities under any arbitrary operations $M_{|i_k}^{A'_k}$, the process matrix $W^\mathbf{A'}$ must be positive semidefinite $W^\mathbf{A'}\geq 0$  and belong to a nontrivial subspace defined by the constraints~\cite{araujo1,oreshkov16,wechs}:
\begin{equation}\label{eq:subspacepm}
    {}_{\prod_{k\in\mathcal{K}}[1-A_k^{O}]\, A_{\mathcal{N}\backslash\mathcal{K}}^{IO}}\, W_{\bm{w}} = 0, \qquad \forall\, \mathcal{K} \subseteq \mathcal{N},\; \mathcal{K} \neq \emptyset.
\end{equation}

with the the trace-and-replace notations
\begin{eqnarray}
    _{[1-X]}W &=& W - _XW, \\
    \label{eq:TeR2} _XW &=& \frac{\id^{X}}{d_X}\otimes \Tr_X W.
\end{eqnarray}

By Theorem~4 of \cite{baumeler22}, any process function $\w: \mathcal{A} \to \mathcal{X}$ admits a unique encoding as  a process matrix:
\begin{align}\label{eq:classicalPM}
    W^{\mathbf{A}'}_{\w}=\sum_{\a} \bigotimes_k \ketbra{a_k}{a_k}^{A_{k}^{O'}}\otimes\ketbra{w_k(\a_{\backslash k})}{w_k(\a_{\backslash k})}^{A_{k}^{I'}}.
\end{align}

Setting $|\I_k|=|\mathcal{O}_k|=1$, arbitrary local operations $f_k:\mathcal{X}_k\rightarrow \mathcal{A}_k$ and $g_{j\neq k}:\mathcal{X}_j\rightarrow \mathcal{A}_j$ encode as quantum channels 
\begin{align}
M_{k}^{A'_k}&=\sum_{x_k\in\X_k}\ketbra{x_k}{x_k}^{A_{k}^{I'}}\otimes
\ketbra{f(x_{k})}{f(x_{k})}^{A_{k}^{O'}},\\
M_{j}^{A'_j}&=  \sum_{x_j\in\X_j}\ketbra{x_j}{x_j}^{A_{j}^{I'}}\otimes
\ketbra{g(x_{j})}{g(x_{j})}^{A_{j}^{O'}}.
\end{align}

\medskip

Let us consider now that each party $k$ performs a single local quantum operation 
\[
M_k: \mathcal{L}(\HS^{A^{I'}_k}) \otimes \mathcal{L}(\HS^{I_k}) \rightarrow \mathcal{L}(\HS^{A^{O'}_k}) \times \mathcal{A}_k 
\]

In comparison with Eq.\eqref{eq:instru}, the classical input register $\mathcal{I}_k$ a quantum one $\mathcal{L}(\HS^{I_k}) $ as  in Eq.\eqref{eq:localpovm}, and we denote the outcome register $\mathcal{O}_k\equiv\A_k$ as well, such that $M_k:=(M_{a_k}^{I_kA'_k})_{a_k}$. Within the process matrix framework, the correlations established by the parties --assuming that they share an auxiliary quantum state $\rho\in\mathcal{L}(\HS^{\mathbf{I}})$-- are then given by the probabilities
\begin{align}
    P(\a|\rho)=\Tr\left[
\big(\bigotimes_{k=1}^n M_{a_k}^{I_kA'_k}\big)^T\big(
\rho^\mathbf{I}\otimes W^\mathbf{A'}\big)
\right],\label{eq:probapm}
\end{align}
where $W\in\mathcal{L}(\HS^{\mathbf{A'}})$ is the process matrix. 

Eq.~\eqref{eq:probapm} can be rewritten as

\begin{align}
P(\a|\rho)=\Tr\left[(E_{\a}^{\mathbf{I}})^T\rho^\mathbf{I}\right]
\end{align}
with 
\begin{align}
E_{\a}^{\mathbf{I}} = M_{\a}^{\mathbf{I} \mathbf{A}'} * W^{\mathbf{A}'}
\label{eq:dpovmelmt}
\end{align}
 with $M_{\a}^{\mathbf{I}\mathbf{A}'}= \bigotimes_{k=1}^n M_{a_k}^{I_kA_k'}$. 

 \medskip

 We now show that the family $(E^{\mathbf{I}}_{\a})_{\a}$ defines an effective, ``distributed” measurement~\cite{supic17,hoban18}. While this follows intuitively from the fact that process matrices always generate valid probabilities under arbitrary local operations, we nevertheless present an explicit proof for completeness and as a sanity check.

\begin{theorem}\label{theo:PF_POVM}
Let $W^{\mathbf{A}'} \in \L(\HS^{\mathbf{A}'})$ 
be a $n-$partite process matrix, and let $(M_{\a}^{\mathbf{I}\mathbf{A}'})_{\a}$ denote the Choi operators of a distributed quantum instrument with $M_{\a}^{\mathbf{I}\mathbf{A}'}= \bigotimes_{i=1}^n M_{a_i}^{I_iA_i'}$ where  $(M_{a_k}^{I_kA_k'})_{a_k}$ is a local quantum instrument performed by party $k$. Then the set of operators $(E^{\mathbf{I}}_{\a})_{\a}$ with $E^{\mathbf{I}}_{\a}$ given by Eq.~\eqref{eq:dpovmelmt}
    defines a valid POVM.
\begin{proof}

First, note that the process matrix $W^{\mathbf{A}'}$ and each local Choi operator $M_{a_k}^{I_k A_k'}$, for $1 \leq k \leq n$, are positive semidefinite. Since the link product preserves positivity, it follows that $E_{\a}^{\mathbf{I}} \geq 0$ for all outcomes $\a$.

\medskip

The proof therefore reduces to demonstrating that $\sum_{\a} E_{\a}^{\mathbf{I}} = \id^{\mathbf{I}}$, which follows by induction.

\medskip

\emph{Base step.} Theorem~\ref{theo:PF_POVM} was shown for $n=2$ in Ref.~\cite[Appendix B]{dourdent21}. For completeness, let us show that it holds for $n=1$. 
For one party, the constraint Eq.~\eqref{eq:subspacepm} writes as $\!_{[1-A_1^{O'}]}W^{A_1'} = 0$, i.e.
$W^{A_1'} = \!_{A_1^{O'}}W^{A_1'}$. Then,

\begin{eqnarray}
    \sum_{a_1} E_{a_1}^{I_1} &=& \sum_{a_1} M_{a_1}^{I_1 A_1'} * W^{A_1'}\notag \\
    \label{eq:indProof2} &=& \sum_{a_1} M_{a_1}^{I_1 A_1'} * _{A_1^{O'}}\! W^{A_1'} \notag \\
    \label{eq:indProof3}&=& \sum_{a_1} M_{a_1}^{I_1 A_1'} * (\frac{\id^{A_1^{O'}}}{d_{A_1^{O'}}}\otimes \Tr_{A_1^{O'}} W^{A_1'})\notag  \\
\label{eq:indProof4}&=& (\frac{1}{d_{A_1^{O'}}}\Tr_{A_1^{O'}}\sum_{a_1} M_{a_1}^{I_1 A_1'}) * ( \Tr_{A_1^{O'}} W^{A_1'}) \notag  \\
    \label{eq:indProof5}&=& \frac{\id^{I_1A_1^{I'}}}{d_{A_1^{O'}}} * ( \Tr_{A_1^{O'}} W^{A_1'}) \notag \\
    &=& \frac{\id^{I_1}}{d_{A_1^{O'}}} \Tr_{A_1'}[W^{A_1'}] \notag  \\
    \label{eq:indProof6} &=& \id^{I_1},
\end{eqnarray}

where we used the trace-preserving constraint $\id^{A_1^{O'}}*\sum_{a_1}M_{a_1}^{I_1A_1'}=\Tr_{A_1^{O'}}\sum_{a_1}M_{a_1}^{I_1A_1'} = \id^{I_1 A_1^{I'}}$ and the normalization constraint $\Tr_{A_1'}[W^{A_1'}] = d_{A_1^{O'}}$, which follows from the definition of process matrices~\cite{oreshkov1, araujo1}.\\

\emph{Induction step.} Assume that given a $n-$partite process matrix,  $\sum_{\a}E_{\a}^{\mathbf{I}} = \id^{\mathbf{I}}$ holds. A $(n+1)-$ partite process matrix satisfies the constraint $\!_{\prod_{k=1}^{n+1}[1-A_k^{O'}]}W^{\mathbf{A}'} = 0$ \cite{araujo1}, which is equivalent to \begin{equation}\label{eq:nPmReduction} 
    W^{\mathbf{A}'} = \sum_{k=1}^{n+1} (-1)^{k-1} \sum_{1\leq i_1 < \ldots < i_k \leq n+1} \!_{\prod_{j=1}^k A_{i_j}^{O'}} W^{\mathbf{A}'}.
\end{equation}
Therefore, we have:
\begin{widetext}
\begin{eqnarray}
    \sum_{\a} E_{\a}^{\mathbf{I}} &=& \sum_{\a} M_{\a}^{\mathbf{I}\mathbf{A}'} * W^{\mathbf{A}'}\notag \\
    \label{eq:indStepProof2}  &=& \sum_{k=1}^{n+1} (-1)^{k-1} \sum_{1\leq i_1 < \ldots < i_k \leq n+1} \sum_{\a} M_{\a}^{\mathbf{I}\mathbf{A}'} *\!_{\prod_{j=1}^k A_{i_j}^{O'}} W^{\mathbf{A}'} \notag\\
    \label{eq:indStepProof4} &=& \sum_{k=1}^{n+1} (-1)^{k-1} \sum_{1\leq i_1 < \ldots < i_k \leq n+1}    \sum_{\a} M_{\a}^{\mathbf{I}\mathbf{A}'} *\Bigl(\id^{\prod_{j=1}^k A_{i_j}^{O'}}\otimes \frac{\Tr_{\prod_{j=1}^k A_{i_j}^{O'}} W^{\mathbf{A}'}}{\prod_{j=1}^k d_{A_{i_j}^{O'}}}  \Bigr)\notag \\
    \label{eq:indStepProof5}  &=& \sum_{k=1}^{n+1} (-1)^{k-1} \sum_{1\leq i_1 < \ldots < i_k \leq n+1} \id^{\prod_{j=1}^k I_{i_j}A^{I'}_{i_j}} * \Biggl(  \sum_{\a \backslash \prod_{j=1}^k a_{i_j}} M_{\a \backslash \prod_{j=1}^k a_{i_j}}^{\mathbf{I}\mathbf{A}' \backslash \prod_{j=1}^k I_{i_j} A_{i_j}'}  * \frac{\Tr_{\prod_{j=1}^k A_{i_j}^{O'}} W^{\mathbf{A}'}}{\prod_{j=1}^k d_{A_{i_j}^{O'}}}\Biggr) \notag\\
    \label{eq:indStepProof6} &=& \sum_{k=1}^{n+1} (-1)^{k-1} \sum_{1\leq i_1 < \ldots < i_k \leq n+1} \id^{\prod_{j=1}^k I_{i_j}} *\Biggl(  \sum_{\a \backslash \prod_{j=1}^k a_{i_j}} M_{\a \backslash \prod_{j=1}^k a_{i_j}}^{\mathbf{I}\mathbf{A}' \backslash \prod_{j=1}^k I_{i_j} A_{i_j}'}  * \frac{\Tr_{\prod_{j=1}^k A'_{i_j}} W^{\mathbf{A}'}}{\prod_{j=1}^k d_{A_{i_j}^{O'}}}\Biggr)\\
    \label{eq:indStepProof7} &=& \sum_{k=1}^{n+1} (-1)^{k-1} \sum_{1\leq i_1 < \ldots < i_k \leq n+1} \id^{\mathbf{I}} \\
    &=& \id^{\mathbf{I}},
\end{eqnarray}
\end{widetext}

where ~\eqref{eq:indStepProof7} relies on the fact that, for any $k$, $\Tr_{\prod_{j=1}^k A_{i_j}} W^{\mathbf{A}'}/\prod_{j=1}^k d_{A_{i_j}^{O'}}$ is 
 a valid $(n+1-k)-$partite process matrix. Consequently, by the induction hypothesis, the operators appearing inside the parentheses in~\eqref{eq:indStepProof6} are valid D-POVM elements, whose sum is the identity, concluding the proof.
\end{proof}
\end{theorem}

Now, let us consider a $n-$partite process function $\w: \mathcal{A} \to \mathcal{X}$. It can be uniquely encoded in a process matrix of the form Eq.\eqref{eq:classicalPM}.
Hence, Theorem \ref{theo:PF_POVM} guarantees that given any process function $\w$ and collection of quantum instruments $(M_{\a}^{\mathbf{I}\mathbf{A}'})_{\a}$, the family $(E_{\a}^{\mathbf{I}})_{\a}$ of operators $E_{\a}^{\mathbf{I}} = M_{\a}^{\mathbf{I} \mathbf{A}'} * W^{\mathbf{A}'}_{\w}$ defines a valid D-POVM.\\

Consider now the scenario in which each party performs a quantum instrument that proceeds as follows. Upon receiving the classical input state $\ket{x_k}^{I_k^{I'}}$ from the process $W_{\w}^{\mathbf{A}'}$, party $k$ performs a projective measurement $(M_{a_k}^{I_k})_{a_k}$ on their quantum input space $\L(\HS^{I_k})$, with  local projectors of the form Eq.\eqref{eq:localpvmelmt}, i.e.\ $M_{a_k}^{I_k}=\ketbra{(a_k|x_k)}{(a_k|x_k)}^{I_k}$. The measurement outcome is then encoded into the classical output state $\ket{a_k}^{I_k^{O'}}$, which is returned to the process. Therefore, each local quantum instrument element can be written
\begin{equation}
M_{a_k}^{I_kA_k'}=\sum_{x_k} M_{a_k|x_k}^{I_k}\otimes \ketbra{x_k}{x_k}^{I_k^{I'}}\otimes\ketbra{a_k}{a_k}^{I_k^{O'}}.
\end{equation}

\medskip

 In order to prove that $(E_{\a}^{\mathbf{I}})_{\a}$ is a D-PVM, it suffices to demonstrate its idempotence, i.e.~for all $\a$, $(E_{\a}^{\mathbf{I}})^2 = E_{\a}^{\mathbf{I}}$.

In turn,
\begin{eqnarray}
    \nonumber E_{\a}^{\mathbf{I}} &=& \sum_{\tilde{\a},\x} \bigotimes_{k=1}^n M_{a_k|x_k}^{I_k} \Tr\Bigl[ |x_k\rangle \langle x_k \ketbra{w_k(\tilde{\a}_{\backslash k})}{w_k(\tilde{\a}_{\backslash k})}^{A_{k}^{I'}}\\
    &\otimes& |a_k\rangle\langle a_k \ketbra{\tilde{a}_k}{\tilde{a}_k}^{A_{k}^{O'}}\Bigr] \\
    &=& \sum_{\tilde{\a},\x} \bigotimes_{k=1}^n \delta_{a_k,\tilde{a}_k} \delta_{x_k, w_k(\tilde{\a}_{\backslash k})} M_{a_k|x_k}^{I_k} \\
    &=& \bigotimes_{k=1}^n M_{a_k|w_k(\a_{\backslash k})}^{I_k}.
\end{eqnarray}
Since, for each $k$, $M_{a_k|w_k(\a_{\backslash k})}^{I_k}$ is a projector, the operator $E_{\a}^{\mathbf{I}}$ is idempotent, which completes the proof.

\end{document}